\documentclass[1pt]{elsarticle}

\usepackage{hyperref}
\usepackage{amsmath,bm} 
\graphicspath{{img/}}
\usepackage[center]{subfigure}

\usepackage{xcolor} 
\usepackage{hyperref} 
\hypersetup{
  colorlinks,
  citecolor=Violet,
  linkcolor=Green,
  urlcolor=Blue
  }
  
\definecolor{changescolor}{RGB}{0, 0, 0}
  
\newcommand\annotation[1]{\textcolor{changescolor}{#1}}

\journal{Journal of Neural Networks}

\AtBeginDocument{%
\renewcommand{\figurename}{Fig.}
\renewcommand{\tablename}{Tab.}
\newcommand{\fig}[1]{\figurename~\ref{fig:#1}}
\newcommand{\tab}[1]{\tablename~\ref{table:#1}}

\newcommand{\eq}[1]{Eq.~(\ref{eq:#1})}
\newcommand{\refsec}[1]{Section~\ref{sec:#1}}
}

\begin{document}

\begin{frontmatter}

\title{A Review on Neural Network Models of Schizophrenia and Autism Spectrum Disorder\corref{cor2}}

\author[add1]{Pablo Lanillos\corref{cor1}}
\ead{p.lanillos@tum.de}
\author[add1]{Daniel Oliva\corref{cor1}}
\ead{daniel.oliva@tum.de}
\author[add2]{Anja Philippsen\corref{cor1}}
\ead{anja@ircn.jp}
\author[add3]{\\Yuichi Yamashita}
\author[add2]{Yukie Nagai}
\ead{nagai.yukie@mail.u-tokyo.ac.jp}
\author[add1]{Gordon Cheng}

\address[add1]{Institute for Cognitive Systems, Technical University of Munich, \\Arcisstra{\ss}e 21, Munich, Germany}
\address[add2]{International Research Center for Neurointelligence, The University of Tokyo, \\7-3-1 Hongo, Bunkyo-ku, Tokyo, Japan
}
\address[add3]{Department of Functional Brain Research, National Center of Neurology and Psychiatry, 4-1-1 Ogawa-Higashi, Kodaira, Tokyo, Japan}

\cortext[cor1]{-authors contributed equally}
\cortext[cor2]{Some figures are under copyright.}
 



\begin{abstract}
This survey presents the most relevant neural network models of autism spectrum disorder and schizophrenia, from the first connectionist models to recent deep network architectures. We analyzed and compared the most representative symptoms with its neural model counterpart, detailing the alteration introduced in the network that generates each of the symptoms, and identifying their strengths and weaknesses. We additionally cross-compared Bayesian and free-energy approaches, as they are widely applied to modeling psychiatric disorders and share basic mechanisms with neural networks. Models of schizophrenia mainly focused on hallucinations and delusional thoughts using neural dysconnections or inhibitory imbalance as the predominating alteration. Models of autism rather focused on perceptual difficulties, mainly excessive attention to environment details, implemented as excessive inhibitory connections or increased sensory precision. We found an excessive tight view of the psychopathologies around one specific and simplified effect, usually constrained to the technical idiosyncrasy of the used network architecture. Recent theories and evidence on sensorimotor integration and body perception combined with modern neural network architectures could offer a broader and novel spectrum to approach these psychopathologies. \annotation{This review emphasizes the power of artificial neural networks for modeling some symptoms of neurological disorders but also calls for further developing these techniques in the field of computational psychiatry.}

\end{abstract}

\begin{keyword}
\texttt{Neural Networks \sep Schizophrenia \sep Autism Spectrum Disorder \sep Computational Psychiatry \sep Predictive Coding}
\end{keyword}

\end{frontmatter}

\tableofcontents

\section{Introduction}

\begin{figure}[hbpt!]
\centering
\subfigure[]{\includegraphics[width=0.45\columnwidth, height=90px]{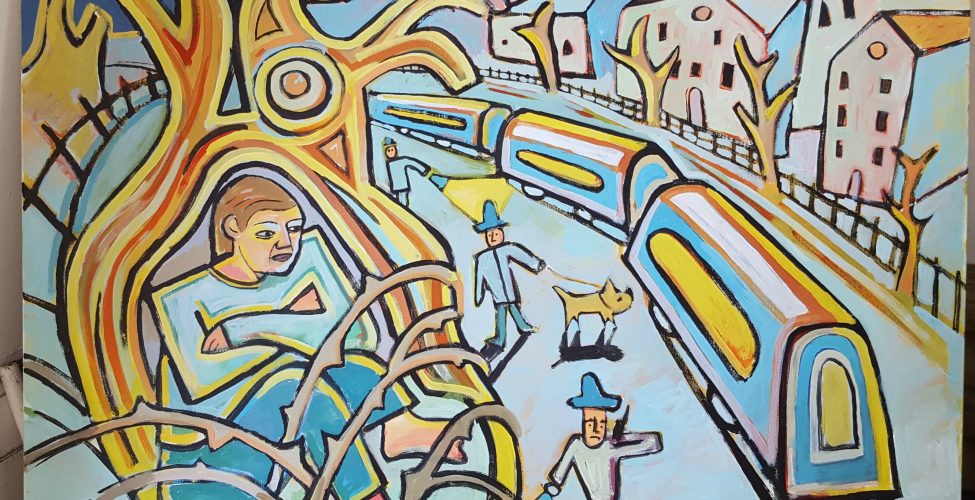} 
\label{fig:painting:schizo}}
\subfigure[]{\includegraphics[width=0.45\columnwidth, height=90px]{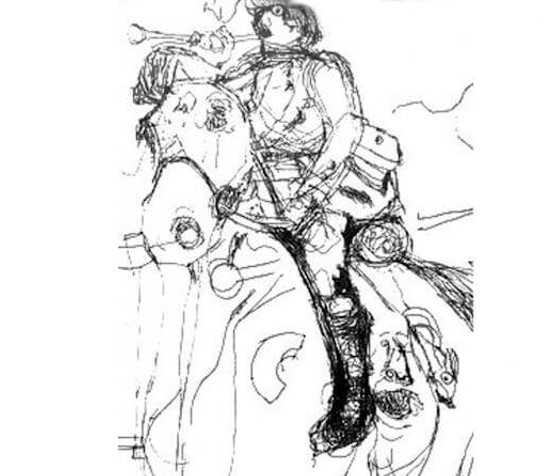} 
\label{fig:painting:asd}}
\caption{Artistic pieces representing different perceptions of the world. (a) Hunted, \textcopyright  2019 Henry Cockburn, a SZ diagnosed artist. (b) Drawing by Nadia Chomyn at the age of 5, a gifted ASD diagnosed child, reprinted from~\cite{selfe2012nadia}, \textcopyright 2012 Lorna Selfe.}
\label{fig:painting}
\end{figure}

In the world, there is a prevalence of schizophrenia (SZ) that ranges between four and seven per 1000 individuals (between three and five million people) \cite{saha2005systematic} and a prevalence of Autism Spectrum Disorder (ASD) that ranges between six and 16 per 1000 children (between 1 of 150 and 1 of 59 children) \cite{baio2018prevalence}. SZ and ASD have in common that they both cause deficits in social interaction and are characterized by perceptual peculiarities. While ASD has its onset in early childhood, SZ is typically diagnosed in adults, although in very rare cases, appears during development \cite{rapoport2009autism}.
Similar neural bases have been observed for both disorders \cite{pinkham2008neural}, which has even led to the suggestion that some SZ cases might be part of the autism spectrum \cite{king2011schizophrenia}.
In fact, there are similarities such that both pathologies show atypical sensorimotor integration and perceptual interpretation.
However, there are also striking differences between these disorders. A common symptom of SZ is the occurrence of hallucinations or delusions, in contrast to ASD which is characterized by atypical non-verbal communication and emotional reciprocity. Furthermore, a few savant syndrome cases were reported in ASD individuals with extraordinary skills like painting \cite{treffert2009savant}.
\fig{painting} depicts, in an artistic way, the reality perceived by two individuals in the spectrum of these disorders.

For both disorders, neurological, genetic and environmental factors have been suggested, but to date the actual causes and underlying cognitive processes remain unclear.
A major challenge in diagnosis is their heterogeneity and non-specificity.
Heterogeneity means that symptoms, prognosis and treatment responses vary significantly between different subjects.
Non-specificity expresses that a single biological basis can be underlying different phenotypes (multifinality) and different biological bases can result in a single phenotype (equifinality). Non-specificity, as a biological abnormality related to a psychiatric disorder, can be found in many other neurological disorders \cite{cross2013identification,redish2016computational}.

Computational modeling of psychopathologies or \textit{Computational Psychiatry} is one of the potential key players \cite{wang2014computational, montague2012computational,redish2016computational} to tackle heterogeneity and non-specificity, and to better understand the cognitive processes underlying these disorders.
Eventually, computational models might help to obtain a deeper understanding of theoretical models, generate new hypothesis or even suggest new treatments.
There are different levels of descriptions or units of analysis to study these disorders, which encompass from genes to molecules, to cells, to circuits, to physiology, and then to behaviour. \textit{``Computational Psychiatry provides some of the tools to link these levels"}  \cite{adams2016computational}.

In particular, neural network models serve, due to their analogy to biological neurons, as a tool to test and generate hypotheses on possible neurological causes \cite{huys2011computational}. Artificial neural networks cannot only be useful from the data-driven point of view (e.g., fitting a model to fMRI\footnote{fMRI: functional magnetic resonance imaging} data), but can also be used as a simplified model of the human brain to replicate and predict human behavior and to investigate which modifications in the connectionist models cause a specific alteration in the behavior.

\subsection{Artificial neural network modeling of psychopathologies}

Artificial Neural Networks (ANNs or NNs) were first introduced in the 1950's as an attempt to provide a computational model of the inner processes of the human brain \cite{rosenblatt1958perceptron}. Nevertheless, their potential was not fully unraveled until the last decades because of limited computational power and data shortage \cite{schmidhuber2015deep}.
Due to the inspiration from biological processes of our brain and their connectionist nature, these technologies have also opened a door to new research fields that combine disciplines, such as neuroscience and psychology with artificial intelligence and robotics. Within the field of cognitive neuroscience, neural networks are already used as a tool for getting insights into the complex structures of our brain and gaining a better understanding of how learning, memory or visual perception might work on a neural level \cite{crick1983function, spitzer1995neurocomputational}.

In the late 80's and early 90's, neural networks were used for the first time related to psychiatry, trying to imitate psychological disorders \cite{hoffman1987computer, cohen1992context}. Early efforts in compiling ANN models for cognitive disorders can be found in \cite{reggia1996neural} and in \cite{gustafsson2004neural}, in particular, for autism. Due to immense advances in computational power, 20 years later, computational modeling using ANNs and deep learning is becoming a powerful asset to aid the investigation of this type of disorders.
The challenge is to translate findings from behavioral or neurological studies at different levels of description in a coherent way into a mathematical connectionist model.

ANN models can process a vast amount of information, cope with non-linearities in the data, and the structure of ANNs makes it possible to systematically test which parameter modifications cause effects similar to the symptoms of psychiatric disorders. Furthermore, these ANN models and their alterations may be directly implemented in artificial agents (e.g., robots) filling the last level: comparing the behavior of such agents with behaviors observed in patients \cite{pfeifer2006body,cheng2007cb}. In this way, existing hypotheses from neuroscience and psychology could be tested, and new hypotheses on potential causes could be formulated.

\subsection{Purpose and content overview}
This historical review aims at serving as a reference for computational neuroscience, robotics, psychology and psychiatry researchers interested in modeling psychopathologies with neural networks. This work extends general computational modeling reviews \cite{reggia1996neural, gustafsson2004neural, anticevic2015bridging, valton2017comprehensive, moustafa2017neurocomputational} by focusing on neural network models for SZ and ASD with detailed explanation of the alterations on a neural level and their associated symptoms, including their technical architectures as well as their mathematical formulation. For completeness, we also included Bayesian and predictive processing models due to their similarities to ANNs and their relevance inside the neuroscience community. Actually, conceptually, ANN and Bayesian models often take similar approaches to model psychiatric disorders (see \refsec{bayesianSchizo} and \refsec{bayesianASD}).



We start in \refsec{pathologies} with an introduction to the mentioned disorders, listing their main characteristics and symptoms based on the latest Diagnostic and Statistical Manual of Mental Disorders (DSM-5) descriptions.


For readability and due to the heterogeneity of the reviewed methods, in \refsec{discussion}, we first summarize and discuss the main modeling approaches and hypotheses which are referenced in the literature. Afterwards, \refsec{modelsschizo} and \refsec{modelsasd} present a comprehensive review of models of SZ and ASD, respectively, organized by the type of modeling approach. To help the reader, we summarized the content of \refsec{modelsschizo} and \refsec{modelsasd} into two tables: \tab{schizophrenia} (page~\pageref{table:schizophrenia}) for SZ and \tab{autism} (page~\pageref{table:autism}) for ASD. 
Finally, in \refsec{new} we discuss the reviewed works and compile recommendations for future research on ANNs for computational psychiatry, in particular for ASD and SZ.

\section{Pathologies and their symptoms}
\label{sec:pathologies}

SZ and ASD are disorders that change the way we perceive and act in the world. Atypicalities in perception and in cognitive process cause difficulties in connecting with the world, in particular for social interaction.
Since the first reports of autistic symptoms \cite{kanner1943autistic}, both conditions have been closely related. Before ASD was recognized as a separate disorder, subjects with ASD were often diagnosed as schizophrenic instead \cite{kanner1943autistic}. Also nowadays, these two pathologies remain strongly connected as both are associated with atypicalities in sensory processing and information processing, and due to their strong heritability \cite{daniels2008parental, aukes2008finding, sandin2017heritability}. 

\subsection{Schizophrenia}

SZ is a serious psychiatric disorder that affects a person's feelings, social behavior and perception of reality. Its biological causes are still unknown, but genetic and environmental factors, i.e., prenatal stress, traumatic experiences or drug use, can be key factors for the development of this disorder. Its symptoms are usually divided into positive symptoms and negative symptoms \cite{sims1988symptoms}. \annotation{Positive symptoms correspond to the presence of abnormal functions, for instance, hallucinations and delusions.} Negative symptoms, corresponding to decreased function, are a lack of the normal function such as diminished emotional expression. Positive symptoms are more apparent and generally respond better to medication. Negative symptoms are more subtle and less responsive to pharmacological treatment. Below some of the most characteristic symptoms of SZ taken from the DSM-5 \cite{american2013diagnostic} are listed. 


\bigskip

\noindent Positive symptoms:
\begin{enumerate}
\item \textit{Delusions}: have convinced beliefs that are not real, and cannot be changed despite clear evidence.
\item \textit{Hallucinations}: perceive things that do not exist as real, without an external stimulus. 
\item \textit{Disorganized thinking}: difficulty to keep track of thoughts, drift between unrelated ideas during speech.
\item \textit{Disorganized or abnormal movements}: difficulties to perform goal-directed tasks, catatonic (stopping movement in unconventional posture) or stereotyped (repetitive) movements.
\end{enumerate}

\noindent Negative symptoms:
\begin{enumerate}
\item \textit{Diminished emotional expression}: reduced expression of emotions through speech, facial expressions or movements.
\item \textit{Avolition}: lack of interests, inaction.
\item \textit{Alogia}: diminished speech output.
\item \textit{Anhedonia}: diminished ability to experience pleasure.
\item \textit{Asociality}: lack of interest in social interaction.
\end{enumerate}

Multiple reports have also associated \textit{self-other disturbances} to SZ. This means that schizophrenic patients can perceive own and external actions or feelings, but may have problems differentiating them. This could be part of the explanation for auditory hallucinations and struggles during social interaction. Van der Weiden and colleagues published an extensive review \cite{van2015self} on possible causes for this disorder. Finally, in more severe cases, motor disorders have been reported \cite{morrens2006stereotypy}, such as stereotypical and catatonic behavior.

SZ is investigated by many researchers because of its prevalence and its devastating effects on patients, which can have life-changing consequences on the patient's relationships and social situation. Moreover, its close relation with the inner workings of self-perception and self-other distinction, raises the interest of researchers from multiple areas such as psychology, neuroscience, cognitive science and even developmental robotics.

\subsection{Autism spectrum disorder}

ASD is a prevalent developmental disorder that has a behavior-based diagnosis due to its still unclear biological causes. It was first introduced in the 1940s by Kanner \cite{kanner1943autistic}, who presented the cases of eleven children ``whose condition [differed] so markedly and uniquely from anything reported so far'', some of them being previously diagnosed as schizophrenic. Actually, the term \textit{autistic} was originally used for describing symptoms in schizophrenic patients. This kind of disorder mainly affects individual's social interaction, communication, interests and motor abilities. It is often referred to as a heterogeneous group (spectrum) of disorders, as individuals typically show distinct combinations of symptoms with varying severity. Nevertheless, there are some characteristic attributes that are commonly associated with ASD, which we have listed from the DSM-5 \cite{american2013diagnostic}. 

\bigskip

\noindent Deficits in social communication and interaction:
\begin{enumerate}
\item \textit{Impairment in socio-emotional reciprocity}: struggle to share common interests and emotions, reduced response or interest in social interaction,
\item \textit{Deficits in non-verbal communication}: problems integrating verbal and nonverbal communication, and using and understanding gestures or facial expressions,
\item \textit{Problems to maintain relationships}: problems or absence of interest in understanding relationships and adjusting behavior.
\end{enumerate}

\noindent Abnormal behavior patterns, interests or activities:
\begin{enumerate}
\item \textit{Stereotyped movements or behavior}: repetitive motor movements or speech,
\item \textit{Attention to sameness}: adherence to routines, distress because of small changes,
\item \textit{Fixated and restricted interests}: strong attachment to certain objects, activities or topics,
\item \textit{Hyper- or hyporeactivity to sensory input}: indifference to pain, repulsive response to certain sounds or textures, visual fascination.
\end{enumerate}

\annotation{Deficits in social interaction are often the most obvious symptoms of ASD. Hence, for a long time, ASD was mainly considered as a disorder of \textit{theory of mind}, suggesting that individuals with ASD are characterized by absence or weakening of their ability to reason about the beliefs and mental states of others in social contexts \cite{baron1997mindblindness}. Actually, early identification of individuals with ASD has focused on non-verbal communication interaction, mainly observing attention and gaze behaviours using standardized tests, such as the Autism Diagnostic Observation Schedule (ADOS) \cite{lord2012autism}. Whereas this explanation could account for a vast amount of symptoms that become obvious in development and socialization of children with ASD, it was mainly criticized due to its failure to explain similarly prominent non-social symptoms such as restricted interests, desire for sameness or excellent performance in specific areas.}

An alternative was suggested in the 90's with the \textit{weak central coherence} theory \cite{frith1994autism, happe2006weak}. It sees the underlying causes of ASD in the perceptual domain, namely in difficulties to integrate low-level information with higher-level constructs. This ``inability to integrate pieces of information into coherent wholes (central coherence)'', stated in \cite{frith2003autism}, could offer explanations for the aforementioned deficits and also be extended to an explanation of social deficits.
An even broader view is provided by the Bayesian brain hypothesis which suggests general deficits in the processing of predictions and sensory information, and can be applied to non-visual perception as well as motor abilities.

ASD is thought to be caused by genetic disorders and environmental factors and evidence points at high heritability \cite{sandin2017heritability}. Furthermore, recent studies, using a computer model of the human fetus, have also highlighted the importance of intrauterine embodied interaction on the development of the human brain and in particular cortical representation of body parts \cite{yamada2016embodied}. Some authors have suggested that preterm infants might have a higher risk of enduring such developmental disorders. 

\section{Modeling approaches and hypotheses}
\label{sec:discussion}


ASD and SZ are among the psychiatric disorders which are most commonly investigated using computational modeling. A reason might be the unclear underlying cognitive mechanisms of these disorders which computational models might help to unravel. The studies we discuss in this review often take similar approaches for modeling ASD and SZ. In fact, these two disorders share certain symptoms, such as deficits in social communication and motor impairments manifesting as decreased response or repetitive and stereotyped movements. \annotation{Although, perceptual atypicalities in both disorders are usually differentiated in that SZ involves perceptual experiences that occur without an external stimulus (e.g., hallucinations) whereas ASD is more typically characterized by hypersensitivity to certain stimuli from the environment, there is some overlap. For instance, hypersensitivity can be also found in SZ patients \cite{robbins1993experiences}. Furthermore, both disorders present less sensitivity to some visual illusions \cite{happe1996studying, notredame2014visual}. Despite of all these similarities, it is still under debate how these two disorders relate to each other \cite{wood2017autism}}.


In computational modeling, similarities between modeling approaches are not primarily motivated by the similarities in symptoms. In fact, studies modeling SZ focused mainly on delusions and hallucinations which are not predominant in ASD. Similarities, instead, can be found in the suggested biological causes and in the type of altered neural network parameters.

There are three main biological causes that are commonly employed in computational models: neural dysconnections\footnote{Note that \textit{dis}connection usually refers to a lack of connection whereas \textit{dys}connection describes atypical connectivity which might include decreased as well as increased connectivity.}, imbalance of excitation and inhibition, and alterations of the precision of predictions or sensory information.











\subsection{Dysconnection hypotheses}

Especially for SZ, one of the most discussed theories is the idea of functional disconnections \cite{friston1998disconnection, lynall2010functional}. The main motivation is that SZ cannot be explained by an impairment of a single brain region, but only by a (decreased) interaction between multiple brain regions \cite{friston1998disconnection}.
Disconnections or underconnectivity are also discussed as a potentional cause of ASD \cite{frith2004autism, just2004cortical, anderson2010decreased}, but more recent evidence also points at increased connectivity \cite{keown2013local, supekar2013brain} or a distortion of patterns of functional connectivity \cite{hahamy2015idiosyncratic}.

In the discussed studies for SZ, dysconnection is primarily implemented by an increased pruning of synapses \cite{hoffman1989cortical, hoffman1997synaptic, hoffman2011using}.
Such a pruning is a normal developmental process between adolescence and early adulthood \cite{huttenlocher1979synaptic}.
Computational models using Hopfield networks \cite{hoffman1989cortical} or feed-forward networks \cite{hoffman1997synaptic, hoffman2011using} demonstrate that too strong pruning can cause fragmented recall or the recall of new patterns, which can be related to the symptom of hallucinations in SZ.

Notably, the SZ symptoms replicated with connection pruning focus solely on hallucinations or delusions and might not be appropriate for modeling ASD.
In fact, in a biological context, it might be more appropriate to disturb connections between neurons instead of simply cutting them.
This idea was followed by Yamashita and Tani \cite{yamashita2012spontaneous} who induced noise between different hierarchies of neurons (suggested by \cite{friston1995schizophrenia}).
They demonstrated in a robotic experiment that this leads to the emergence of inflexible, repetitive motor behavior similar to catatonic symptoms in SZ. This motor behaviour could also be present in ASD.


Just a single study focused on dysconnection in ASD. Park and colleagues \cite{park2019macroscopic, ichinose2017local} showed, using a spiking neural network, that local over-connectivity, especially locally in the prefrontal cortex \cite{courchesne2005frontal}, can account for the emergence of aberrant frequency patterns of neural connections in patients with ASD.


\subsection{Excitation/inhibition imbalance}

An excitation/inhibition (E/I) imbalance is among the most commonly referenced biological evidence for SZ as well as for ASD \cite{rubenstein2003model, sun2012impaired, snijders2013atypical, canitano2017autism}.
E/I imbalance was found in many neurobiological studies on SZ and ASD.
Although it is not clear how exactly E/I imbalance translates to changes in cognition and behavior \cite{canitano2017autism}, it seems to be linked to core symptoms of both disorders such as hallucinations \cite{jardri2016hallucinations} and social interaction deficits \cite{yizhar2011neocortical}.

An unanswered question is also of which quality this imbalance is. A recent review of studies regarding ASD found evidence for increased inhibition as well as for increased excitation \cite{dickinson2016measuring}.
Conflicting results in various brain regions might arise by differences in measurements and their reliability. The most commonly used mechanisms are magnetic resonance spectroscopy which allows to measure the cortical levels of glutamate or GABA, measurements of gamma-band activity (which is hypothesized to be connected to inhibition) or the analysis of the number of glutamate or GABA receptors in post-mortem studies \cite{dickinson2016measuring}.
Another possible interpretation of these conflicting results is that both, increases and decreases, in inhibition and excitation are present in ASD.
This hypothesis was put forward by Nagai et al. \cite{nagai2015influence}, suggesting that both impairments share a common underlying mechanism. Their model could show that increased inhibition and increased excitation can simulate the local or global processing bias of ASD, respectively.

Furthermore, Gustafsson \cite{gustafsson1997inadequate} also connected E/I imbalance to the local processing style of ASD. He implemented increased inhibition in a self-organizing map, in particular, stronger inhibition in the surrounding of receptive fields which led to over-discrimination.

For SZ, although E/I imbalance is commonly associated to SZ in the literature, only the approach from Jardri et al. \cite{jardri2013circular} explored E/I imbalance as a modeling mechanism.
In their model, a stronger excitation or insufficient inhibition caused circular belief propagation: bottom-up and top-down information \annotation{are confused with each other which might cause hallucinations and delusions (see page~\pageref{sec:circinf})}.
This model was recently supported by some experimental evidence \cite{jardri2017experimental}.



\subsection{Hypo-prior theory and aberrant precision account}

The increasing popularity of the Bayesian view on the brain in recent years resulted in a trend of explaining psychiatric disorders as a cause of the failure of correctly integrating perceived low-level sensory information (bottom-up information) with high-level prior expectations (top-down information). These approaches are inspired by diminished susceptibility of subjects with psychiatric disorders to visual illusions \cite{notredame2014visual} and the well-known symptom of hypersensitivity to certain stimuli (e.g., \cite{lucker2013auditory}).


Problems in the integration of top-down and bottom-up information can be explained by an inadequate estimation of the precision of these signals. A decreased precision of the prior causes a weaker reliance on predictions and, hence, a relatively stronger reliance on sensory input. This so-called hypo-prior theory was first suggested by Pellicano and Burr for ASD in 2012 \cite{pellicano2012world}. Similarly, an increased precision of the bottom-up signal can account for the same consequences \cite{lawson2014aberrant}. Despite some initial evidence in favor of an overrating of sensory information \cite{karvelis2018autistic}, it cannot be decided to date which of these theories is more compelling than the other. Possibly, both contribute to the observed phenomena.

For both, ASD and SZ, typically a weaker influence of predictions and a higher influence of sensory information is suggested \cite{pellicano2012world, lawson2014aberrant, karvelis2018autistic}. Lawson and colleagues substantiated aberrant precision for ASD by basing it on hierarchical predictive coding. \annotation{They argued that both hypo-priors and increased sensory noise might influence the perception on different levels of the cortical hierarchy}, leaving open both hypotheses.
In an endeavor to clarify how such theories differ for ASD and SZ, Karvelis et al. \cite{karvelis2018autistic} recently investigated how healthy individuals, scored for traits of ASD and SZ, use prior information in a visual motion perception task. \annotation{ASD traits were associated with increased sensory precision, whereas SZ traits did not correlate}. 

However, it might be intuitively plausible that also an overrating of top-down information can account for the occurrence of hallucinations \cite{powers2016hallucinations}.
In a recent review, Sterzer et al. \cite{sterzer2018predictive} noticed that too strong as well as too weak priors explain psychosis. They suggested that the way that priors are processed might differ depending on the sensory modality or the hierarchical level of processing, yielding inconsistent theories and findings.

In line with this idea, computational models for ASD often suggest that an impairment might be present in both extremes \cite{idei2017reduced, philippsen2018understanding}. In \cite{idei2017reduced}, repetitive movement could be replicated by an aberrant estimation of sensory precision, leading to inflexible behavior, either due to sameness of intentional states (increased sensory variance) or due to high error signals and misrecognition (decreased sensory variance). Similarly, \cite{philippsen2018understanding} suggests that too strong as well as too weak reliance on the sensory signal may impair the internal representation of recurrent neural networks. Thus, for SZ as well as for ASD, too strong as well as too weak reliance on priors or sensory information seem to be valid modeling approaches.




\subsection{Alternative modeling approaches}

There are alternative theories used in the discussed computational models.
Synaptic gain, for instance, has been evaluated for SZ \cite{cohen1992context} as well as for ASD \cite{dovgopoly2013connectionist}. In fact, a reduction of synaptic gain might be related to reduced precision of prior beliefs as discussed in \cite{adams2018bayesian}.

\annotation{Less biologically inspired approaches can also be found in the literature and focus more on replicating behavioural data using known engineering techniques in ANN. For instance, deficits in generalization capabilities are modeled in neural networks by modifying the number of neurons \cite{cohen1994artificial}, changing the training time \cite{dovgopoly2013connectionist} or introducing regularization factors \cite{dovgopoly2013connectionist, ahmadi2017bridging}.}

\section{ANN models of schizophrenia}

\label{sec:modelsschizo}

\begin{table*}[!hbtp]
\caption{Overview of neural network models of schizophrenia}
\centering
\resizebox{\textwidth}{!}{
\begin{tabular}{ | l | p{4cm} | p{5cm} | p{5cm} | p{5cm} |}
        \hline
        \textbf{Model type} & \textbf{Paper} & \textbf{Disorder Characteristic} & \textbf{Biological Evidence} & \textbf{Approach} \\ \hline
        Hopfield Networks & R. E. Hoffman, T. H. McGlashan (1987)\cite{hoffman1987computer} & Delusions, sense of mind being controlled by outside force & - & Storing of an excessive number of memories (memory overload)\\ \hline
        & R. E. Hoffman, T. H. McGlashan (1989)\cite{hoffman1989cortical} & Hallucinations, delusions, sense of mind being controlled by outside force & Reduced connectivity in prefrontal cortex and other regions & Excessive connection pruning\\\hline
        & D. Horn, E. Ruppin (1995) \cite{horn1995compensatory} & Delusions and hallucinations & Reactive synaptic regeneration in frontal cortex  & Weakening of external input projections, increase of internal projections and noise levels, additional Hebbian component\\\hline
        
        Feed-forward NNs & J. D. Cohen, D. Servan-Schreiber (1992) \cite{cohen1992context} & Disturbances of attention, representation of context & Abnormal dopamine activity in prefrontal cortex & Reduction of activation function gain in context-neurons\\ \hline
        & R. E. Hoffman, T. H. McGlashan (1997) \cite{hoffman1997synaptic} & Auditory hallucinations & Reduced connectivity in prefrontal cortex and other regions & Excessive connection pruning\\ \hline
        & R. E. Hoffman et al. (2011) \cite{hoffman2011using} & Delusionary story reconstuction & Abnormal dopamine activity, cortical disconnections & Increased BP learning rates, excessive connection pruning in working memory\\ \hline
        
        Predictive processing & Adams et al. (2013) \cite{adams2013computational} & Delusions and hallucinations, abnormal smooth pursuit eye movement & Abnormal neuromodulation of superficial pyramidal cells in high hierarchical levels & Abnormal precision computation in the free energy minimization scheme\\\hline
        Circular inference & Jardri and Denéve (2013) \cite{jardri2013circular} & Hallucinations and delusions & Disruption in the neural excitatory to inhibitory balance & Increased excitation / reduced inhibition in belief propagation \\\hline
        
        Recurrent NNs & Y. Yamashita, J. Tani (2012) \cite{yamashita2012spontaneous} & Disturbance of self, feeling of being controlled by outside force, disorganized movements & Disconnectivities in hierachical networks of prefrontal and posterior brain regions  & Noise between context neuron hierarchies in MTRNN\\\hline
        \end{tabular}
        }
\label{table:schizophrenia}
\end{table*}


In the following section, we present a comprehensive description of the most important ANN models of SZ. The majority of approaches focuses on positive symptoms of SZ, such as hallucinations and delusional behavior, e.g., \cite{horn1995compensatory} and \cite{hoffman1997synaptic}. Nevertheless, there have been also approaches targeting other symptoms, for instance attention characteristics \cite{cohen1992context} and movement disorders \cite{yamashita2012spontaneous}.
An overview of the most important models is presented in \tab{schizophrenia}.





\subsection{Hopfield networks: memory }

\subsubsection{Memory overload}

\label{SchizoHoffman87}

In 1987, Ralph E. Hoffman, professor of psychiatry from Yale, presented the earliest neural network model of SZ \cite{hoffman1987computer}, inspired by the suggestions of \cite{crick1983function}, who explored the function of dreams using a neural network model. Hoffman tried to explain the causes of schizophrenic and maniac disorders with simulations using a Hopfield Network, an associative memory ANN that is usually employed to simulate the inner functioning of human memory \cite{hopfield1982neural} and to store binary memory patterns. It is a recurrent neural network that converges to fixed-point attractors. As a learning mechanism, the famous Hebbian rule, ``cells that fire together wire together'', is applied. In other words, connections between neurons that get activated with temporal causality are increased \cite{hebb1949organization}. In order to model SZ, the author inspected the behavior of the network attractors after storing an increasing number of binary memories.



Results showed that by \textbf{increasing the number of binary memory patterns stored}, the network reaches ``parasitic'' states that do not correspond to previously stored memories. With higher numbers of memories or decreased storage capacity, the network's internal energy minima, that correspond to the stored memories, might influence each other and create additional deep minima (attractors) that do not correspond to any previously learned pattern. These minima might influence either only the information processing course (mind being controlled by outside force) or lead to convergence to ``parasitic states'', which are compared to hallucinations and delusional thoughts. This study did not use biological evidence to support its main thesis that SZ might be caused by memory overload and only compared behavioral observations.
However, this model served as a stepping stone for a successor model (see \refsec{SchizoHoffmanHop}).

\subsubsection{Memory model with disconnections}

\label{sec:SchizoHoffmanHop}

\begin{figure}
    \centering
        \includegraphics[scale=0.45]{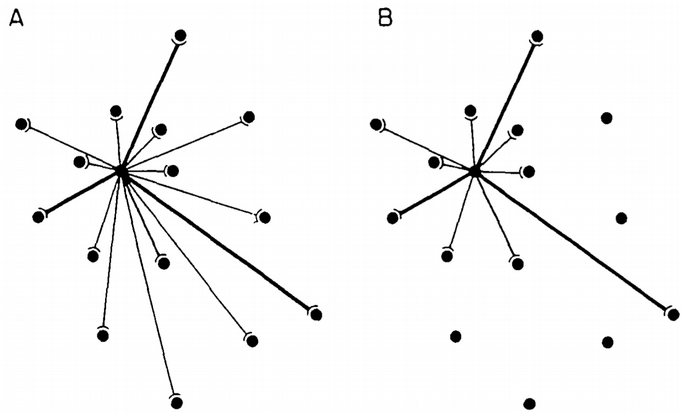}
    \caption{Pruning rule used for the Hopfield Network in \cite{hoffman1989cortical}. The connections are pruned depending on the connection weight and the distance between the connected neurons. A: Connections before pruning. B: Connections after pruning. Reprinted from \cite{hoffman1989cortical}.}
    \label{fig:HoffmanHopfieldPruning}
\end{figure}

Observations that show diminished metabolism in the prefrontal cortex (hypofrontality) of individuals with SZ led to the theory that excessive synaptic pruning might be the reason for the appearance of SZ between adolescence and early adulthood \cite{feinberg1982schizophrenia, keshavan1994schizophrenia}.
A decline in synaptic density is a normal developmental process \cite{huttenlocher1979synaptic, huttenlocher1982synaptogenesis} which might have gone too far in the case of SZ. In 1989, Hoffman and Dobscha used a Hopfield network, \annotation{arranged as a 2D grid}, as a content-addressable memory to retrieve previously stored memories giving a similar input \cite{hoffman1989cortical}. A ``neural Darwinism'' principle was applied, which is a \textbf{pruning rule that erases connections depending on their weights and length \annotation{(proximity of neurons in the grid)}}. The concrete pruning rule is shown in \eq{PruningRule}, with $|T_{xy}|$ being the weight of the connection between neurons in coordinates $(x,y)$ and $(i,j)$, and $\hat{p}$ the pruning coefficient. The pruning coefficient determines the number of connections which are discarded. \fig{HoffmanHopfieldPruning} illustrates a possible scenario for this pruning process.

\begin{equation}
    |T_{xy}| = \boldsymbol{\hat{p}} \cdot [(i-x)^2 + (j-y)^2]^{0.5}
    \label{eq:PruningRule}
\end{equation}

For a moderate level of pruning, the network is still able to perform the memory-retrieval task, but for connection reductions of 80\% the network shows fragmented retrieval. This fragmentation was compared to thought disorders observed in SZ, which lead to incoherence, attention deficits or the feeling that one's mind is being controlled by an outside force. Furthermore, sometimes over-pruned areas converged to patterns not included in any of the stored memories. These were denominated as ``parasitic foci''. The authors compared these to hallucinations in SZ because they contained decodable information that does not belong to any stored memory. Occasionally, these parasitic regions extended on a larger area and persisted independently of the input, which was compared to delusional thoughts observed in patients.

\subsubsection{Memory model hippocampal region}
\label{sec:SchizoRuppin}

In 1995, Horn and Ruppin \cite{horn1995compensatory, ruppin1996pathogenesis} also introduced a Hopfield-based network to replicate the positive symptoms of SZ. This model was based on the hypothesis by J. R. Stevens \cite{stevens1992abnormal} that schizophrenic symptoms might be caused by ``reactive anomalous sprouting and synaptic reorganization taking place at the frontal lobes, subsequent to the degeneration of
temporal neurons projecting at these areas''. The hypothesis takes into account observations that showed atrophic changes in the temporal lobe, and at the same time increased dendritic branching in the frontal lobe of a significant number of schizophrenic patients. Essentially, the idea is that degenerations in temporal lobe regions that are connected to the frontal lobe regions might produce a compensatory reaction in that area, namely increased receptor bindings (frontal lobe connections) and anomalous dendritic sprouting (increased influence from other cortical areas).

\annotation{The work by Hoffman explained in the previous section suggested that hallucinations should always appear in combination with memory problems in patients because pruning clearly affects the network's memory retrieval performance. However, this is not always the case in patients.}
Following the hypothesis from Stevens, the model described in \cite{horn1995compensatory} would make hallucinations and intact memory capabilities compatible.

The model used in this paper was a Hopfield network taken from \cite{tsodyks1988associative, tsodyks1988enhanced}, which is more appropriate for the storage of correlated patterns. This network is used for a pattern retrieval and recovery task, which means that in its original functionality, it receives an external input pattern and outputs the previously learned pattern that corresponds to it, given that a similar one was learned before. 

Defining the connection strength (weight) between neuron $i$ and $j$ as $W_{ij}$, the learning rule is:

\begin{align}
W_{ij_{\:new}} &= \boldsymbol{c} \; W_{ij_{\:old}} \; , \; (c > 1) \\
    W_{ij} &=  \frac{c_0}{N}\sum_{\mu=1}^{M}(\xi_i^\mu - p)(\xi_j^\mu - p)
    \label{eq:InWeight}
\end{align}
where $\boldsymbol{c}$ is the internal projection parameter with value always $>1$. \eq{InWeight} describes the initial configuration of the network weights, with $c_0 = 1$, $p$ being the probability that a memory pattern is chosen to be 1, and $\xi_i^\mu$ one of the $M = \alpha N$ memory patterns. 


The input of each neuron $i$ at time step $t$ is expressed as:
\begin{equation}
   h_i(t) = \sum_{j}W_{ij}S_j(t-1)+ \boldsymbol{e} \cdot \xi_i^1
   \label{eq:Presyn}
\end{equation}
where $e$ is the network input parameter with value $1$ in normal conditions, which weights the incoming memory pattern, and $S_j$ is the neuron output defined by a sigmoid function with noise level $T$ and a fixed uniform threshold of all $N$ neurons $\theta$:

\begin{equation}
   \centering
   S_i(t) =
   \begin{cases}
      1, & \text{with probability } \frac{1}{1 + \exp(-(h_i(t) - \theta)/\boldsymbol{T})} \\
      0, & \text{otherwise}
   \end{cases}
   \label{eq:State}
\end{equation}

In order to simulate degenerated temporal lobe projections to the frontal lobe, the input is scaled down by decreasing parameter $e<1$ in \eq{Presyn}. In order to model increased receptor bindings and dendritic sprouting the parameter $\boldsymbol{c}$ in \eq{InWeight} and noise level $\boldsymbol{T}$ in \eq{State} are increased. The parameter $c$ scales the internal weights of the network and $T$ influences the neuron activation.
After performing these modifications, the network is still able to retrieve previously stored memories, but spontaneously converges to certain memories without a specific input stimulus.





\annotation{
An additional Hebbian learning rule during pattern retrieval on a lower time scale is used to account for increased dopamine levels observed in patients with SZ:}

\begin{equation}
   W_{ij}(t) = W_{ij}(t-1) + \frac{\gamma}{N}(\bar{S}_i - p)(\bar{S}_j - p)
   \label{eq:Hebbian}
\end{equation}
\annotation{where $\bar{S}_i$ is a variable that only becomes $1$ if the neuron in question has been active during the last $\tau$ iterations. There are studies that have observed that dopamine activity increases may enhance Hebbian-like activity-dependent synaptic changes in the brain, and a high synaptic modification rate $\gamma$ is used to replicate this effect, as this parameter influences how much the network's weights are changed during learning. This modification is used to imitate high dopamine levels observed in schizophrenia.}


\begin{figure}[hbtp!]
    \centering
        \includegraphics[width=0.8\textwidth]{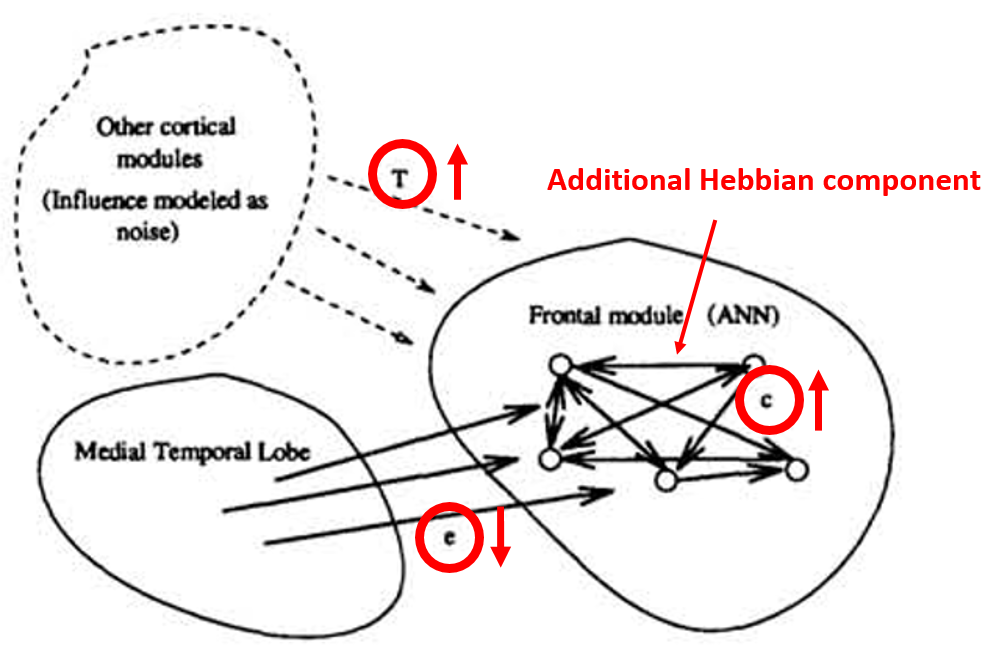}
    \caption{Schematic illustration of the proposed model: An ANN models the frontal module, receiving input from internal connections c, external connections from the medial temporal lobe e and connections T from distant cortical modules modeled as external noise.
    Highlighted in red are the modifications made on the Hopfield network to imitate schizophrenic behavior: Decrease of external input projections, and increase of internal projections and external noise. Adapted from \cite{horn1995compensatory}}.
    \label{fig:RuppinSchizo}
\end{figure}

In total, four network modifications were tested on the presented architecture (\fig{RuppinSchizo}):
(1) \textbf{weakening of the network input parameter $\boldsymbol{e}$}, (2) \textbf{increase of internal projections $\boldsymbol{c}$}, (3) \textbf{increase of noise levels $\boldsymbol{T}$}, and (4) \textbf{additional Hebbian learning rule} (\eq{Hebbian}).

Combining the reactive modifications to a decrease of $e$ (internal connections and external noise) with the described Hebbian rule (even with a small $\gamma$ of 0.0025), the spontaneous retrievals are enhanced and get continuously triggered without a concrete retrieval input. This behavior is compared to long-term hallucinations or delusional beliefs characteristic of schizophrenic patients. This results would also fit with the effect of dopaminergic blocking agents (equivalent to reducing the effect of the Hebbian learning rule), which are used to reduce hallucinations in patients.

\subsection{Feed-forward networks: context and language}


\subsubsection{Attention and context representation}
In 1992 the first model based on feed-forward neural networks was introduced. The psychology professor Jonathan D. Cohen and neuroscientist David Servan-Schreiber \cite{cohen1992context} presented an extensive analysis of a possible explanation for negative symptoms in SZ. More concretely, they focused on disturbances of attention and contextualization problems in schizophrenics, which were for instance reported in \cite{garmezy1977psychology} and \cite{lang1965psychological}. Their main hypothesis was that schizophrenics fail to make an internal representation of context and that an abnormal amount of dopamine in the prefrontal cortex is the main cause (cf. \refsec{SchizoRuppin} as a comparison). The authors refer to previous studies suggesting that the prefrontal cortex is the brain region responsible for maintaining an internal representations of context, and that patients with SZ show dysfunctions and abnormal dopamine levels in this area. In order to test the dopamine-theory of SZ, three experimental tasks were compared to three neural network models, obtaining similar results to empirical observations. They simulated reduced dopamine activity by decreasing the gain of the activation function (the activation function's slope), described by \eq{GainEq}, in the neurons responsible for context representations. In this equation, we used the same nomenclature as in the original paper, where $net$ is the added activation of all incoming connections, $bias$ the neuron bias and $gain$ the parameter that is modified. The mentioned idea of modifying the activation function's gain was based on studies that suggest that high dopamine levels potentiate the neurons' activation (inhibitory and excitatory) in the prefrontal cortex. The modification of the gain has a similar effect because higher gain values increase the activation function's slope, which means that even small neuron input values produce either very low neuron activations (equivalent to inhibitory signals) or high activations (equivalent to excitatory signals).

\begin{equation}
    f(net) = \frac{1}{1 + \exp(\boldsymbol{gain} \cdot net + bias)}
    \label{eq:GainEq}
\end{equation}
 
\begin{figure}[hbpt!]
\centering
\subfigure[Stroop test]{\includegraphics[width=0.45\columnwidth, height=100px]{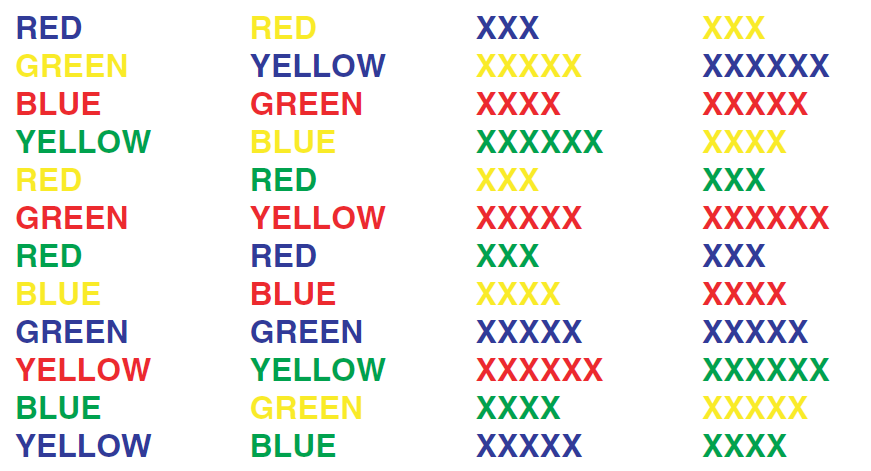} 
\label{fig:stroop:test}}
\subfigure[Network model]{\includegraphics[width=0.45\columnwidth, height=110px]{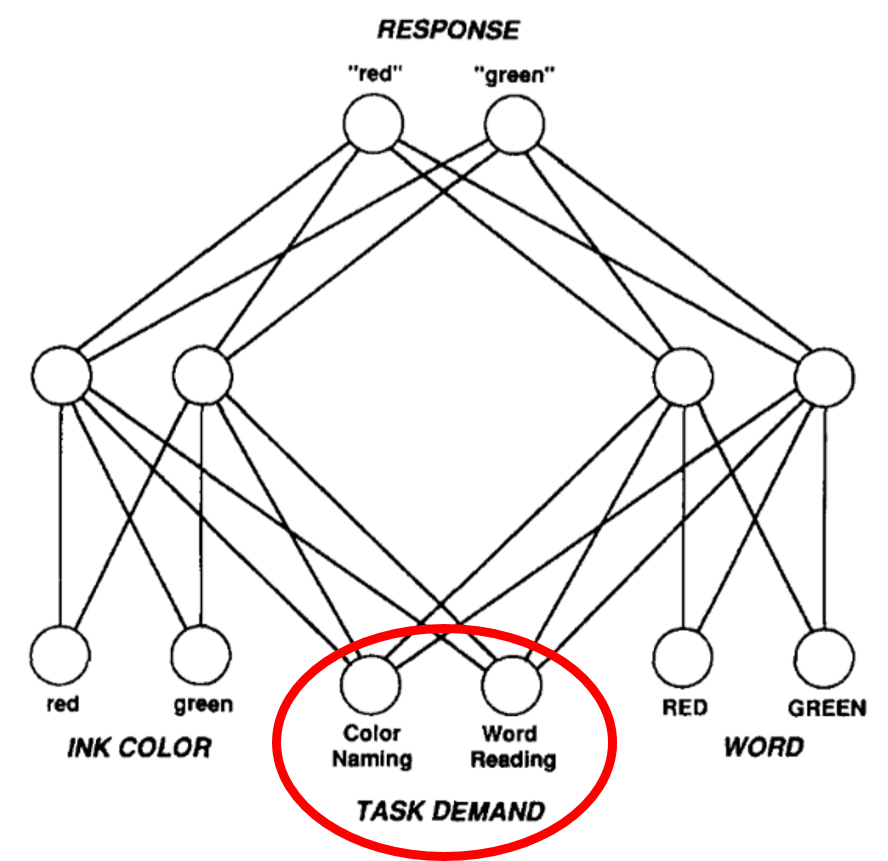}
\label{fig:stroop:model}}
\caption{Attention and context. (a) Stroop card test used for SZ, reprinted from  \cite{henik2004schizophrenia} (b) Neural network model used for the \textit{Stroop} task in \cite{cohen1992context}. Highlighted in red are the neurons with modified gain.}
\label{fig:stroop}
\end{figure}

The first experiment, depicted in \fig{stroop}, was the \textit{Stroop} task \cite{stroop1935studies}, which consists of color words printed in different color inks that are presented to the participants. These words have either congruent stimuli (color and word are the same), conflicting stimuli (color and word contradict each other) or control stimuli (color words printed in black ink or the letters \textit{``XXX"} printed in a certain color). The subjects must then either always name the letter's ink color or the written word. This exercise is used to test the participant's attention capacities, and schizophrenic subjects show overall slower reaction times and perform even worse when conflicting stimuli are shown \cite{henik2004schizophrenia}.
In order to feed the information in the network, the printed word's ink color and meaning were numerically coded. By reducing the gain on the \textit{color naming} and \textit{word reading} units from $1.0$ (normal gain) to $0.6$ they observed a delay in the response time of the network to properly produce a correct answer, similar to what it was observed in schizophrenic diagnosed individuals.

\begin{figure}[!t]
    \centering
    \subfigure[CPT test]{\includegraphics[width=0.48\columnwidth, height=100px]{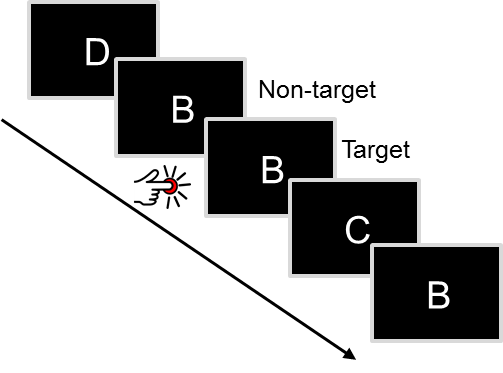}
\label{fig:CPT:test}}
\subfigure[Network model]{\includegraphics[width=0.44\columnwidth, height=110px]{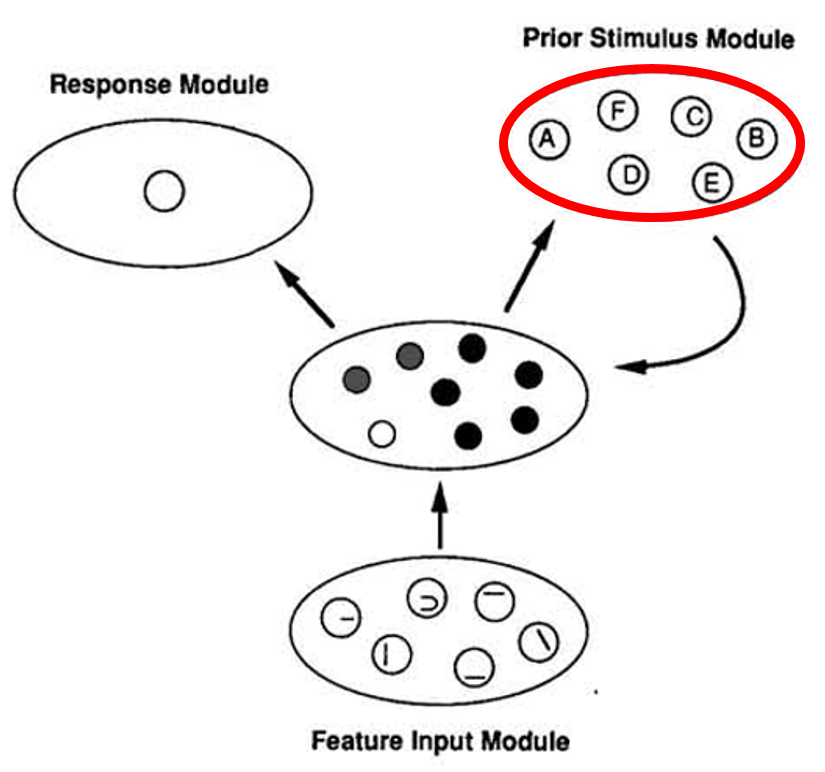} 
\label{fig:CPT:model}}
    \caption{Continuous performance test. (a) Simplified CPT Identical Pair test used (b) Neural network model for the \textit{CPT} adapted from \cite{cohen1992context}. Highlighted in red are the neurons whose gain was decreased to model disturbed processing in the prior stimulus module.}
    \label{fig:CPT}
\end{figure}

The second experiment, shown in \fig{CPT}, implemented the \textit{Continuous Performance Test (CPT)} \cite{rosvold1956continuous} identical pair version \cite{cornblatt1989continuous}. It measures participant's ability to detect repeated pattern of symbols in a longer sequence. Symbols are presented sequentially and the volunteers must detect when the pattern appears consecutively, words or numbers, e.g., ``9903''. In this experiment, schizophrenics usually struggle with the detection of longer patterns where previous symbols need to be taken into account. \textit{Prior stimulus module} neurons were used to save the information about previous sequence symbols. To simulate schizophrenic behavior, the authors \textbf{reduced the gain of the activation-function} of the task context yielding to a higher miss-rate in concordance with schizophrenic empirical observations.

\begin{figure}[t]
    \centering
\subfigure[Task]{\includegraphics[width=0.43\columnwidth, height=110px]{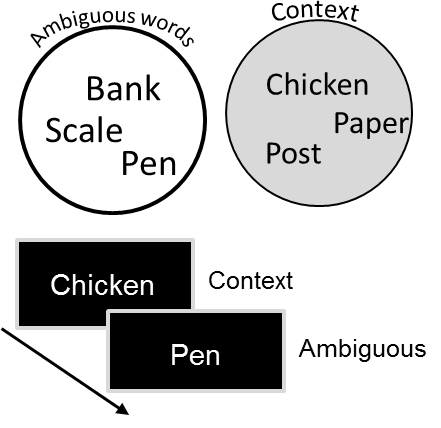}
\label{fig:lexical:test}}
\subfigure[Network model]{\includegraphics[width=0.43\columnwidth, height=120px]{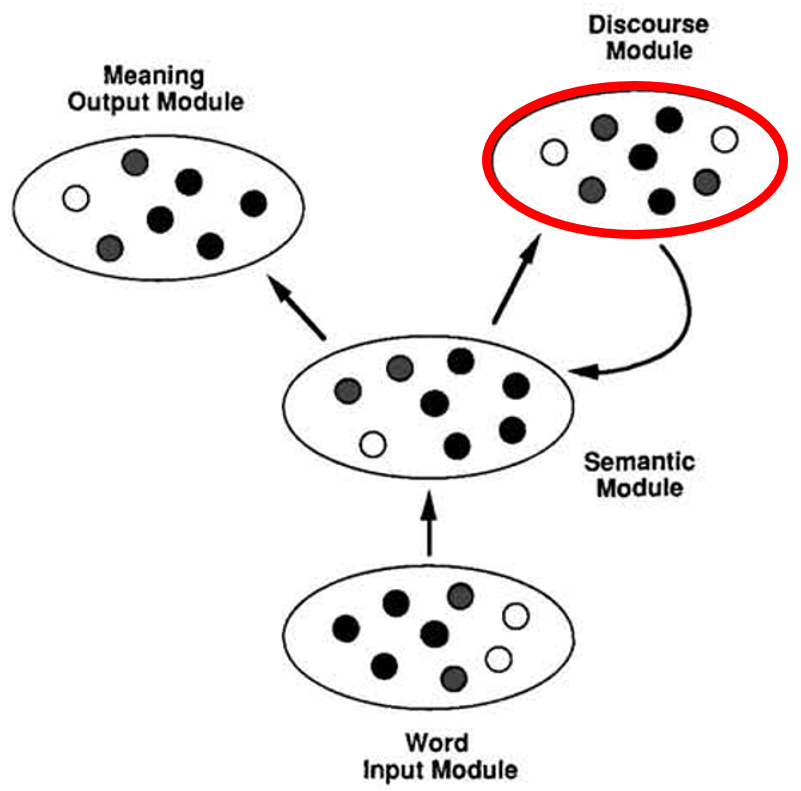} 
\label{fig:lexical:model}}
\caption{Lexical disambiguation. (a) Task with context dependent meaning word. (b) Neural network model reprinted from \cite{cohen1992context}. Highlighted in red are the (context) neurons whose gain was reduced to $0.6$.}
    \label{fig:lexical}
\end{figure}

Finally, a lexical disambiguation task depending on context was modeled based on the original work from Chapman et al. \cite{chapman1964theory} (see \fig{lexical}). Participants had to solve homonym conflicts (words with more than one meaning), taking into account the context of the sentence. In this case, schizophrenics show worse performances when the needed context to resolve ambiguity comes before the word in question. A similar approach than in the CPT experiment was taken: context neurons gain was manually reduced to $0.6$ like in the previous experiments. It resulted in low performance for the schizophrenic model when the sentence context that was needed to interpret the ambiguous word was located at the beginning of the sentence.

\subsubsection{Auditory processing}

\label{SchizoHoffmanFF}



\begin{figure}[hbtp!]
    \centering
        \subfigure[Network model]{\includegraphics[width=0.48\columnwidth, height=100px]{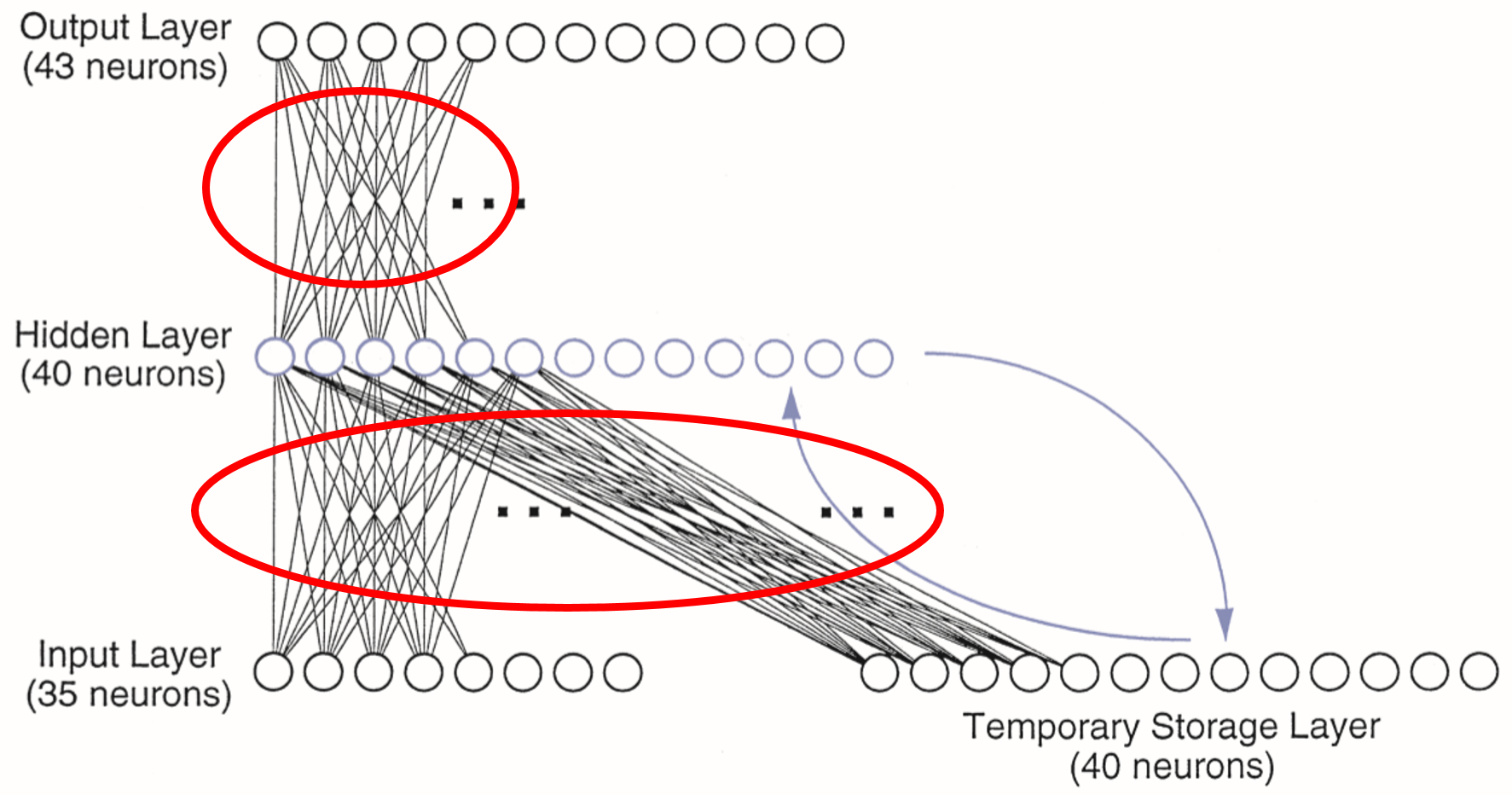} 
\label{fig:HoffmanResults:model}}
    \subfigure[Word detection results]{\includegraphics[width=0.48\columnwidth, height=100px]{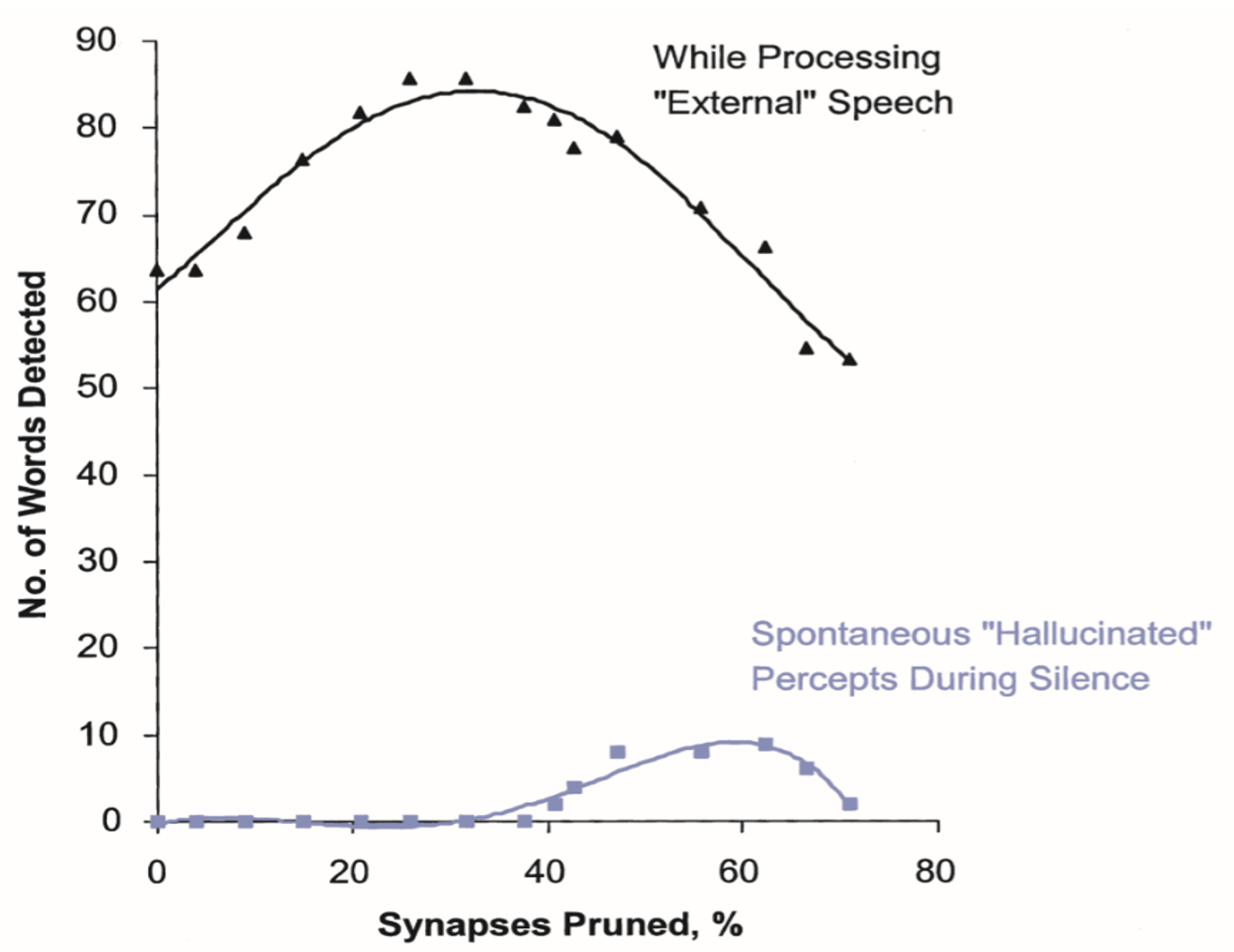} 
\label{fig:HoffmanResults:results}}
    \caption{Auditory hallucinations (a) Neural network model used in \cite{hoffman2001book}. Input of the network are simulated phonetic codes, output are semantic features of the input word. Highlighted in red are the connections the pruning rule was applied on to imitate schizophrenic symptoms. (b) Word detection results depending on connection pruning. Spontaneous detections are observed for excessive pruning. Reprinted from \cite{hoffman2001book} with permission.}
    \label{fig:HoffmanResults}
\end{figure}

During a person's life, the number of neurons in the brain peaks during childhood and then decreases by a 30\% to 40\% in adolescence, which is also the period of time where SZ appears most frequently (adolescence/early adulthood) \cite{huttenlocher1979synaptic}. Based on this observation and post-mortem findings which suggest neural deficits in the schizophrenic's cerebral cortex \cite{keshavan1994schizophrenia,margolis1994programmed}, Hoffman and McGlashan designed a feed-forward neural network capable of translating phonetic inputs into words \cite{hoffman1997synaptic}. This model was inspired by Elman's (1990) model \cite{elman1990finding}. As illustrated in \fig{HoffmanResults:model} it consists of one hidden layer and a \textit{temporal storage layer} that saves a copy of the hidden layer from the previous processing step.



A \textbf{pruning rule} was used to set the value of \textbf{all connections below a certain threshold to zero}. After pruning approximately 30\% of the connections, the word detection capabilities of the used network improved\footnote{Pruning is a bioinspired standard technique for improving generalization of the network. However, nowadays, dropout approaches have gained popularity over pruning.}. However, with excessive pruning the network starts to struggle with detection tasks and shows spontaneous responses during periods without input (shown in \fig{HoffmanResults:results}). This last observation was associated to auditory hallucinations reported in patients with severe SZ. Furthermore, it supported the common theory that auditory hallucinations might be caused by false identification of own inner speech as externally generated.

In posterior tests with healthy patients, schizophrenics with auditory hallucinations showed reduced word detection capabilities compared to schizophrenics without such hallucinations, which fits with the previous simulations. Furthermore, a later review of this paper \cite{hoffman2001book} highlighted that by applying active repetitive transcranial magnetic simulation (active rTMS) on the left temporoparietal cortex, a brain region usually associated to speech perception, hallucinations seem to be reduced. This further supports the hypothesis of a possible correlation between speech-processing disorders and auditory hallucinations.

 \subsubsection{Language processing}
 
 \label{SchizoHoffmanDISCERN}
 
 Another feed-forward model of SZ introduced by R. E. Hoffman and collaborators \cite{hoffman2011using} uses a network called DISCERN \cite{miikkulainen1991natural, miikkulainen1993subsymbolic, grasemann2007subsymbolic} that is able to learn narrative language and reproduce learned content, e.g., learn a story and reproduce it after feeding it with a fraction of the story.
 

Based on previous studies about SZ, eight different network modifications were tested: (1) \textit{Working Memory (WM) disconnections} by pruning of connections with a weight below a certain threshold, (2) \textit{Noise addition in working memory} by adding of Gaussian noise to WM neuron outputs, (3) \textit{WM network gain reduction} by reducing the activation function's gain, (4) \textit{WM neuron bias shifts} by increasing neuron bias and inducing an increased overall activation, (5) \textit{Semantic network distortions} by adding noise to word representations in semantic memory, (6) \textit{Excessive activation semantic networks} by increasing neuron outputs in semantic network, (7) \textit{Increased semantic priming} by blurring semantic network outputs, (8) \textit{Exaggerated prediction-error signaling (hyperlearning)} by increasing back-propagation learning rates.

The resulting network behaviors were compared to empirical results using a goodness-of-fit measure (GOF), which compared factors such as story recall success (successfully retelling story), agent confusions (switching of certain story characters), lexical errors and derailed clauses (false interpretation of certain sentences). The authors concluded that \textbf{(1) WM disconnections} with pruning and \textbf{(8) hyperlearning} best explain real-world data. These results for WM disconnections further reinforce the previously presented theory by Hoffman and McGlashan in \cite{hoffman1997synaptic} that excessive connection pruning during human's adolescence might be one of the causes for this disorder. Moreover, the authors also suggested that over-learning in schizophrenic brains might cause modifications in previously stored memories, which might lead to delusional or erroneous convictions.

\subsection{Bayesian approaches}
\label{sec:bayesianSchizo}

Several important models of psychiatric disorders are based on the idea that the brain uses Bayesian inference as a basic principle. The Bayesian brain hypothesis describes the human brain as a generative model of the world that makes predictions about its environment and adapts its internal model depending on the observation provided by the senses.
For SZ as well as for ASD it is suggested that patients might differ in the way they combine sensory inputs with prior information. The idea was highly influenced by Hermann Helmholtz's work in experimental psychology \cite{von1867handbuch} that dealt with the brain's capacity to process ambiguous sensory information. \annotation{In his words: \textit{``Visual perception is mediated by unconscious inferences"}. }

\begin{figure*}[hbtp!]
	\centering
	\subfigure[Arcimboldo's painting] {\includegraphics[width=0.23\columnwidth, height=80px]{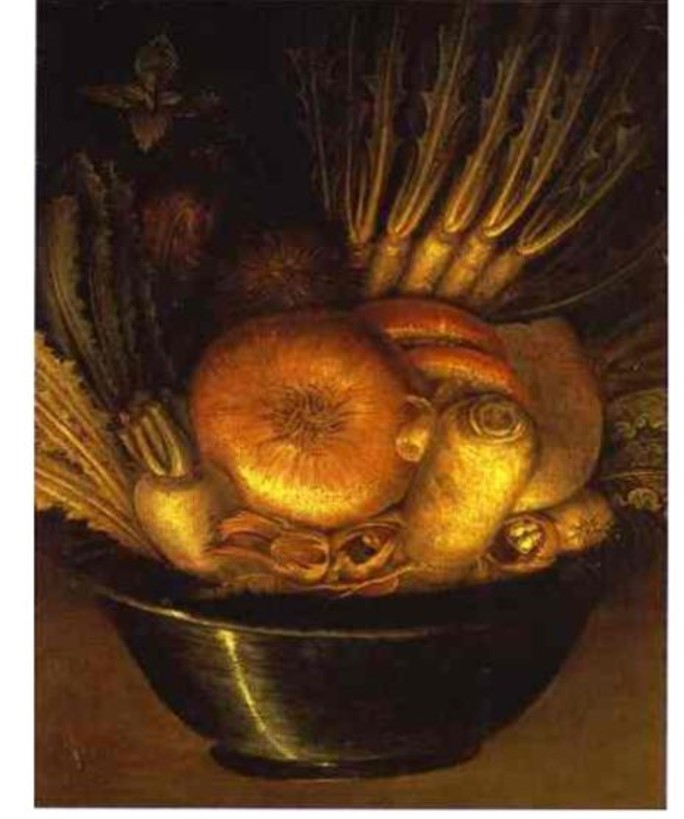} 
		\label{fig:visualillusions:arcimbolo}}
	\subfigure[Tacher's illusion]{\includegraphics[width=0.23\columnwidth, height=80px]{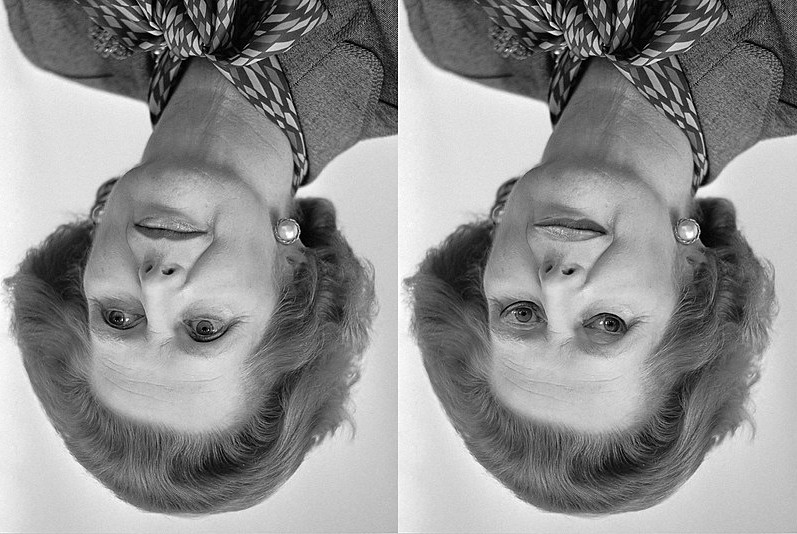} 
		\label{fig:visualillusions:tacher}}
	\subfigure[Ocampo's painting]{\includegraphics[width=0.23\columnwidth, height=80px]{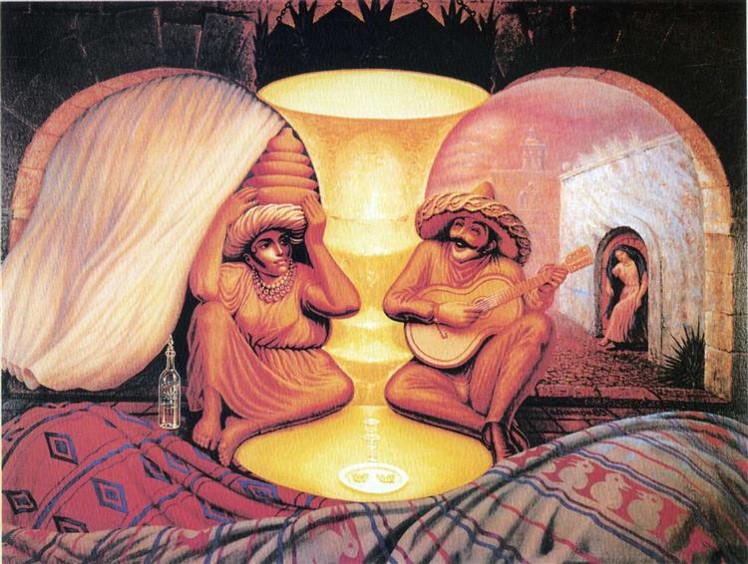} 
		\label{fig:visualillusions:ocampo}}
	\subfigure[Dallenbach's illusion]{\includegraphics[width=0.23\columnwidth, height=80px]{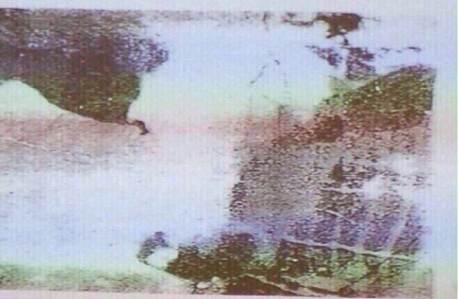} 
		\label{fig:visualillusions:dallenbach}}
		
	\caption{Visual illusions where the brain infers different interpretations depending on the prior information or context. (a) Ortaggi in una ciotola o l'Ortolano. G. \textcopyright Arcimboldo 1590. (b) Tacher illusion \cite{thompson1980margaret}. (c) Forever Allways, \textcopyright Octavio Ocampo 1976. (d) Dallenbach's illusion 1952~\cite{kmd1951puzzle}.}
	\label{fig:visualillusions}
\end{figure*}

\fig{visualillusions} shows puzzle images that stress that perception depends on prior knowledge as well as sensory input. For instance, if we rotate Arcimboldo's painting by 180~degree instead of vegetables we will see a human face with a hat. Tacher's illusion can be broken by also rotating the upside down images, and we will see that both faces are different. In particular, mouth and eyes are inverted. In Ocampo's painting, we can see two old people from a larger distance but two mariachis when viewing the picture from close range. Finally, Dallenbach's illusion shows that even if you know that there is an animal looking at you in the picture, it is impossible to see it until the shape of the cow is highlighted. Afterwards you cannot stop seeing it. In essence, what we perceive not only depends on the raw sensory information, but also on our prior knowledge and predictions we have about the world.

\begin{figure}[hbtp!]
    \centering
        \includegraphics[width=0.9\columnwidth, height=100px]{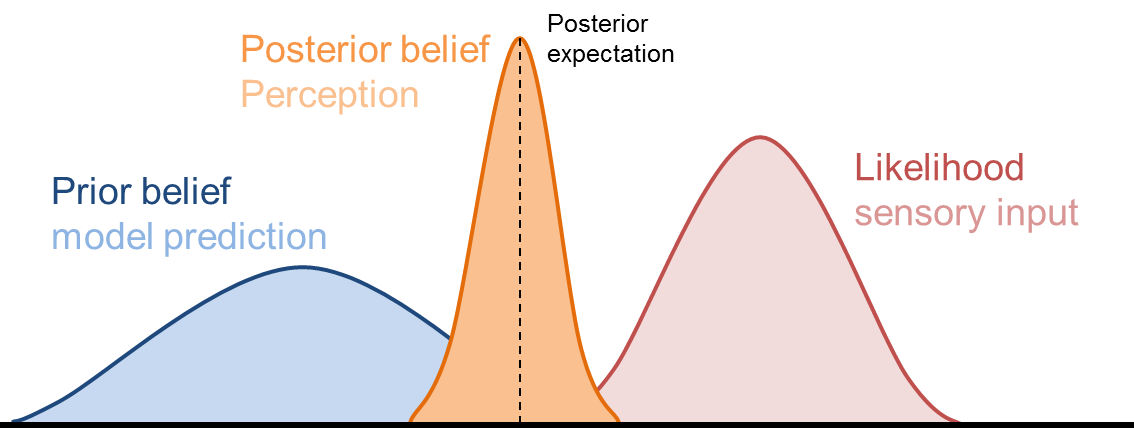}
    \caption{Illustration of Bayesian inference: The posterior belief is generated by inference of  prior belief and sensory evidence. Depending on the variance (precision) of prior and sensory evidence, the posterior belief will be influenced more by one of the previous. Adapted from \cite{adams2013computational}.}
    \label{fig:BayInf}
\end{figure}

\annotation{The classical concept} of Bayesian inference presents perception as computing the posterior belief from the sensory input (likelihood) and from the model prediction (prior belief) depending on their relevance. For instance, in the case of a very imprecise (highly variable) prior, the perception would shift more strongly to the direction of the sensory input. \annotation{\fig{BayInf} illustrates these concepts assuming that the world is one-dimensional and can be described via Gaussian distributions}.

\subsubsection{Free-energy model of schizophrenia}

\label{SchizoFriston}

Friston's free-energy model \cite{friston2010free} describes the brain functionality as a dynamical inference network.
It combined the Helmholtz machine ideas \cite{dayan1995helmholtz} with the hierarchical prediction error message passing \cite{rao1999predictive} and the Bayesian mathematical framework.
Despite not being implemented as an ANN model, we included it in this review because it is considered one of the most relevant models in the computational neuroscience community. \annotation{Furthermore, it serves for comparative purposes with predictive coding neural network implementations of psychiatric disorders \cite{yamashita2008emergence, yamashita2012spontaneous, philippsen2018understanding}.}

Under the free-energy principle, the brain is seen as a prediction machine that progressively constructs an internal model of the world which is constantly improved, based on the received sensory feedback and the resulting prediction error.
Perception (posterior belief) then results from combining the brain's predictions (prior) with the sensory evidence (likelihood) as shown in \fig{BayInf}. If the prior's precision is relatively higher than the precision of sensory evidence, the posterior will be more similar to the prior. In the opposite case, the posterior will be more close to sensory input. Therefore, precision weights the influence of prior and sensory evidence on the posterior belief. 

Mathematically, the internal model is updated by minimizing the negative \textit{free energy} $F$ a lower bound on the KL-divergence that quantifies the difference between the internal belief about the world and reality. 



Assuming that $\vec{\mu}$ are the dynamical internal states of the brain, perception is then  described as the adaption of $\vec{\mu}$ given the sensory observations by minimizing the free energy using the gradient descent method described in \eq{equationInternalStates}:
\begin{align}
\dot{\vec{\mu}}(t) = D \vec{\mu}(t) - \frac{\partial F(\vec{s},\vec{\mu})}{\partial\vec{\mu}} = D\vec{\mu}(t) -\frac{\partial \epsilon}{\partial \vec{\mu}} \Pi \epsilon
\label{eq:equationInternalStates}
\end{align}  
where $D$ is a differential matrix operator that computes the currently expected hidden state, $\epsilon$ is the error between the predicted (sensory) input from the higher layer and the real input (observation) and $\Pi$ \annotation{is the inverse variance (precision) of the information}. For instance, in humans, visual information would typically have higher precision than proprioceptive sensing for body localization \cite{hinz2018drifting}.




Based on these concepts, Adams et al. \cite{adams2013computational} built a computational model of SZ and analysed in three different experiments: \annotation{auditory pattern recognition (using the example of a bird recognizing its own song)}, a object eye-tracking task and a simulation of force-matching illusion. \annotation{One of the core ideas was that a reduction of the precision at higher levels of the cortical hierarchy (i.e., reduced precision of prior beliefs) influenced the responses of the model.}
More concretely, decreases in prior precision (or, for the force-matching illusion, failure to reduce sensory precision) led to struggles in auditory pattern recognition, problems with eye-tracking with occlusion and attribution of agency.
Furthermore, with an additional compensatory decrease of sensory precision (for the force-matching illusion, increase of prior precision), the model showed hallucination-like behavior during the auditory pattern recognition task and difficulties to distinguish self-touch and touch from others in the force-matching illusion.

\begin{figure}[hbtp!]
    \centering
        \includegraphics[width=0.9\columnwidth, height=200px]{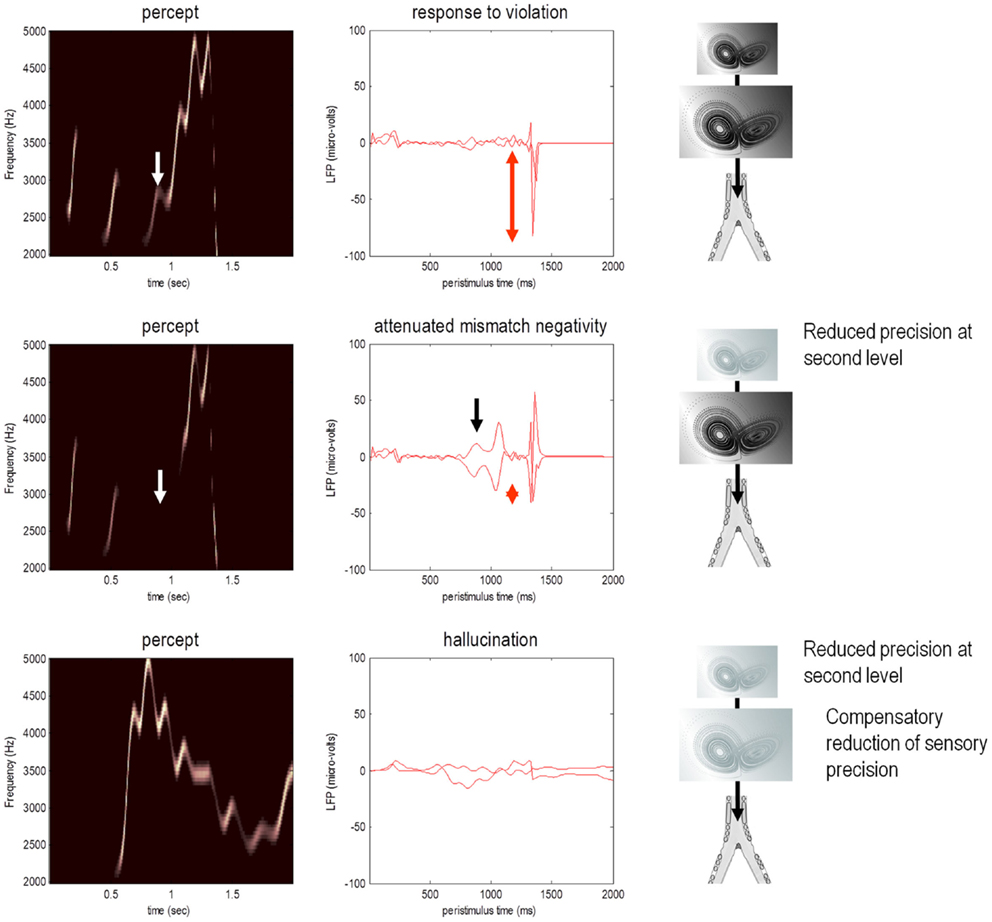}
    \caption{Prediction sonograms of the auditory signal of a birdsong (left), prediction error with respect to stimulus (middle) and used model (right), when last three chirps are omitted. Top row: Unmodified model generates prediction error increases with the first missing chirp, which corresponds to normal behavior. Middle row: With reduced precision at second level the model is unable to predict the third chirp, and the prediction error for missing chirps is reduced. Bottom row: With compensatory sensory precision reduction in first level, there is a complete failure of perceptual inference. Despite the wrong predictions, almost no prediction error is generated due to missing precise sensory information. This behavior is compared to auditory hallucinations. Reprinted from \cite{adams2013computational} with kind permission.}
    \label{fig:FristonBirdsong}
\end{figure}

\annotation{\fig{FristonBirdsong} shows the experiment of auditory pattern recognition of a birdsong, showing how the precision in different cortical levels changes the response to surprising events. The first row describes a normal behavior to surprising events (the belief precision is high). In this case, when a chirp of the bird is omitted, the posterior perception contains an illusory (weakened) response at the point in the signal where sensory input is missing (white arrow at left plot). This effect might correspond to omission-related responses found in electrophysiological recordings of the brain \cite{nordby1994erps}. The middle and bottom rows correspond to abnormal behaviours in line with SZ findings, such as attenuation of omission-related responses and auditory hallucinations respectively.}

\subsubsection{\annotation{Circular inference in} Bayesian graphical models}
\label{sec:circinf}

\annotation{In \cite{jardri2013circular}, Jardri and Den\'eve investigated how excitatory to inhibitory imbalance may relate to psychotic symptoms in schizophrenia, using belief propagation in a hierarchical Bayesian graphical model.
In particular, it is shown that \textbf{a dominance of excitation causes circular belief propagation}: bottom-up sensory information and top-down predictions are reverberated, and therefore, may be confused with each other or taken into account multiple times.
The model can account for the occurrance of erroneous percepts (hallucinations) and fixed false beliefs (delusions) in SZ.}


\annotation{In the graphical model, low hierarchical levels correspond to sensory experience and high levels to top-down predictions.
Messages are passed between nodes in different hierarchical levels from lower to higher levels (bottom-up processing) and from higher to lower levels (top-down processing).
The fact that connections exist in both directions raises an important challenge: to differentiate between \textit{real} sensory information and sensory information which were simply \textit{inferred} from top-down expectations.
The authors suggest that such circular belief propagation in the Bayesian network is avoided if a careful balance between excitation and inhibition is maintained. A disruption of this balance can account for the appearance of schizophrenic symptoms.}

\annotation{Concretely, information between higher and lower levels are exchanged in the form of messages.
For belief propagation, messages are passed recursively until convergence:}
\begin{equation}
M_{ji}^{n+1} =
  \begin{cases}
   W_{ij}(B_i^n - \alpha_d M_{ij}^n)   & \quad \text{if } i \text{ is above } j\\
    W_{ij}(B_i^n - \alpha_c M_{ij}^n)  & \quad \text{if } j \text{ is above } i,
  \end{cases}
\end{equation}
where the term \textit{above} means that the node $i$ is in a higher hierarchical level than $j$.
$M_{ij}^n$ is the message sent from $i$ to $j$ at step $n$.
$W_{ij}$ is the connection strength, and $\alpha_d$ and $\alpha_c$ are the parameters that scale the inhibitory loops in upward and downward direction, respectively.
\annotation{$B_i^n$ is the computed belief expressed as a log-odd ratio\footnote{Log-odd ratio: computed as the log of the ration between the probability that a cause is present and that the cause is absent, thus, values around $0$ describe uncertain states, positive values correspond to belief in presence, negative values to belief in absence.} and updated as:}
\begin{equation}
  B_i^{n+1} = \sum_i{M_{ji}^{n+1}}.
\end{equation}

The authors experimented with the two $\alpha$ parameters in this framework, adjusting them between 1 (normal level of inhibition) and 0 (no inhibition). Simulated results show that equally impaired loops (same $\alpha$ below 1) are still able to arrive at a proper inference. Conversely, with unbalanced impaired upward loops ($\alpha_u < 1$) ``over-estimation of the strength of sensory evidence and an underweighting of the prior" is produced. This is compatible with over-interpretation of sensory evidence and the reduced influence to illusions observed in schizophrenic patients.

The authors recently demonstrated in \cite{jardri2017experimental} that the circular inference model nicely fits decisions of SZ diagnosed patients using the Fisher task as the experimental paradigm. The Fisher task permits the manipulation of the prior and the likelihood allowing comparisons with the Bayesian model predictions. Participants have to decide whether the fish captured comes from the left or the right lake. First, two boxes (left, right) with fish and different sizes are presented (prior): bigger box express higher probability. Secondly, the two lakes (left, right) are presented with fishes inside with two colors (red and black). The proportion of red fishes represent the likelihood of the observation. Finally, participants have to decide if the red fish comes from the left or the right. According to the participant's data and their proposed model, descending and ascending loops correlated with negative and positive SZ symptoms respectively.


\subsection{Recurrent neural networks}

\label{SchizoTani}

In 2012, Yamashita and Tani presented a model of SZ using a recurrent neural network (RNN) \cite{yamashita2008emergence} such as they are commonly used for the recognition and generation of time series. Specifically, in this study, the RNN is applied to the task of sensorimotor sequence learning in a humanoid robot: the robot learns to predict visual information and own motor movements in a scenario where it moves a cube on a surface.

\begin{figure}
    \centering
        \includegraphics[width=0.8\columnwidth, height=180px]{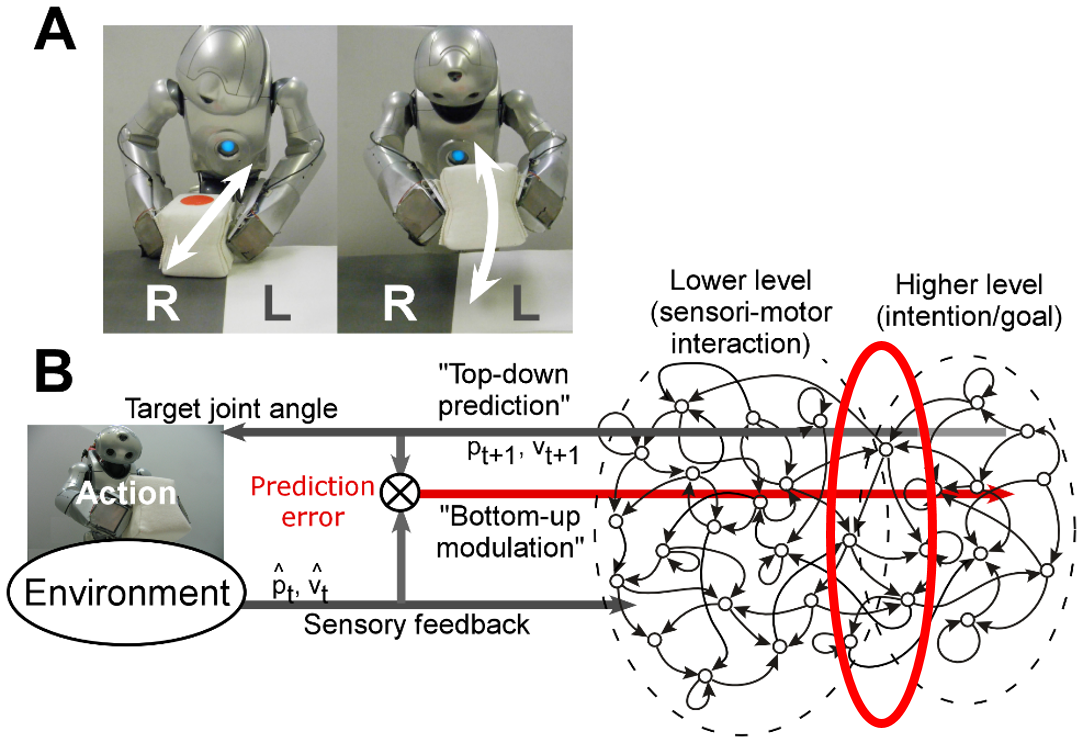}
    \caption{(A) Tasks to be performed by the robot: when the object is on the Right move the object backward and forward, when the object is on the Left move the object up and down. (B) MTRNN network architecture. Highlighted with a red ellipse are the connections between fast and slow context units that are degraded with noise to imitate schizophrenic behavior. Adapted from \cite{yamashita2012spontaneous}.}
    \label{fig:TaniSchizo}
\end{figure}

The type of RNN they used is the Multiple Timescale Recurrent Neural Network (MTRNN), a special type of RNN that mimics the hierarchical structure of biological motor control systems. Human and animal motor movements are commonly suggested to be segmented into so-called ``primitives" \cite{schaal2000nonlinear}. These primitives can then be reused and combined to more complex motor sequences.
The MTRNN contains neurons working at different timescales: fast context neurons (corresponding to the lower level of the hierarchy) learn the motion primitives and slow context units (corresponding to higher, more abstract levels) control the sequence of the primitives (see \fig{TaniSchizo}).
This network is trained to perform prediction error minimization, i.e., to build an internal model of the world following the Bayesian brain idea.
Training the network using the Backpropagation Through Time algorithm (BPTT), the robot learns multiple motions (grasping and moving an object) adapting to different object positions. It is also able to combine these actions into new action sequences by only training the slow context units. The trained network works as a predictor where the sensory input modulates the changes on the slow context units (goals) depending on the error\footnote{There is a strong parallelism between Multiple Timescale RNNs and the hierarchical model proposed by Friston.}.




Equation \ref{MTRNNUpdate} describes the dynamics of each neuron at each layer:
\begin{equation}
    \tau \dot{u}_{i,t} = -u_{i,j} + \sum_{j}w_{i,j} \cdot x_{j,t}.
    \label{MTRNNUpdate}
\end{equation}
In this formula, the membrane potential $u_{i,t}$ of neuron $i$ at time step $t$ is updated with the neural state $x_{j,t}$ of neuron $j$ scaled with the (learnable) connection weights $w_{i,j}$.
The time constant $\tau$ determines the update frequency of the neuron.
A small time constant is used for fast context units, and a large time constant for slow context units.

Schizophrenics can have trouble to distinguish self-generated actions from others' actions and, in severe cases of SZ, patients can even have problems performing movements, and show repetitive or stereotypical behavior \cite{van2015self}.
Based on observations that suggest that SZ may be caused by disconnections in hierarchical brain regions, mainly between prefrontal and posterior regions \cite{friston1995schizophrenia,banyai2011model}, \textbf{uniformly distributed random noise was added in the connections between fast and slow context units} highlighted with the red circle in \fig{TaniSchizo}.
For the evaluation of the model a humanoid robot was used. It had the task of locating an object on a table in front of it and performed different actions depending on the object's position: if the object was located to the right, the robot was supposed to grab the object and move it back and forth three times. Otherwise, if the object was located to the left, the robot had to grab the object and move it up and down three times.

They showed that for a small degree of disconnection (small noise addition) the robot had no problems to perform the mentioned task. Nevertheless, increases of spontaneous prediction error were observed and abnormal state switching appeared in the intention-network (slow units). The authors compared these prediction errors to patient's problems in attribution of agency (when own movements are perceived as being executed by someone else). Schizophrenics might want to perform an action and have an internal prediction of the upcoming proprioceptive and external states. The increases of prediction error could be seen as incongruences between the intended actions and the results, which can give a person the feeling of not being able to control the consequences of its own actions or it may have problems to perceive these actions as self-generated. For more severe disconnections, the humanoid robot clearly struggled to perform the given task and showed disorganized sequences of movements. These observations were compared to more severe cases of SZ, where cataleptic (stopping) and stereotypical (repetitive) behaviors have been observed.



\section{ANN models of autistic spectrum disorder}
\label{sec:modelsasd}
This section describes the most important ANN models of ASD. They focused on the atypical processing style suggested by the weak central coherence theory which could be summarized as excessive attention to detail. They replicated deficits in perception \cite{cohen1994artificial, dovgopoly2013connectionist, gustafsson1997inadequate, nagai2015influence}. Some also addressed atypicalities in memory structure and internal representations \cite{mcclelland2000basis, philippsen2018understanding} and inflexibility in motor behavior \cite{idei2017reduced}. Although most studies suggested connections to social deficits in an indirect way, only one of the models made a direct connection to theory of mind, by modeling weak central coherence on the level of logical reasoning \cite{o2000autism}. 
An overview of the reviewed approaches is given in \tab{autism}.

\begin{table*}[!hbtp]
\caption{Overview of neural network models of ASD}
\centering
\resizebox{\textwidth}{!}{
\begin{tabular}{ | l | p{4cm} | p{5cm} | p{5cm} | p{5cm} |}
        \hline
        \textbf{Model type} & \textbf{Paper} & \textbf{Disorder Characteristic} & \textbf{Biological Evidence} & \textbf{Approach} \\ \hline
        
        Feed-forward and simple recurrent NNs & I. L. Cohen \cite{cohen1994artificial, cohen1998neural} (1994, 1998) & Generalization deficits due to excessive attention to detail & Abnormal neural density in various brain regions & Excessive or reduced number of neurons, increased training duration\\ \hline
        
         & J. L. McClelland (2000) \cite{mcclelland2000basis} & Hyperspecificity of memory concepts & -- & Excessive conjunctive coding\\\hline
         
         & Dovgopoly \& Mercado \cite{dovgopoly2013connectionist} (2013) & Deficits in visual categorization and generalization & Abnormalities in synaptic plasticity & Reduced learning rate, negative weight decay (anti-regularization) \\ \hline

        Self-Organizing Maps & L. Gustafsson (1997) \cite{gustafsson1997inadequate} & Excessive attention to detail & Lateral inhibition enhances sensory perception &  Excessive inhibitory lateral feedback\\\hline
        
         & L. Gustafsson et al. (2004) \cite{gustafsson2004self} & Avoidance of novelty & -- &   Familiarity preference, higher weighting of close data points\\\hline
         
         & G. Noriega (2007) \cite{noriega2007self} & Domain-based hypersensitivity & Early brain overgrowth in children with ASD & Variable (increasing) number of neurons, stronger/weaker attention to stimuli \\\hline

         & G. Noriega (2008) \cite{noriega2008modeling} & Domain-based hypersensitivity & Early brain overgrowth in children with ASD & Propagation delays in neural weight updates \\\hline
         
        Convolutional NN & Y. Nagai et al. (2015) \cite{nagai2015influence} & Local/global processing bias & Excitation/inhibition imbalance & Excitation/inhibition imbalance in visual processing\\\hline
        
        Spiking NNs & J. Park et al. (2019) \cite{park2019macroscopic} & Atypical neural activity: High power in higher frequency bands and decreased signal complexity & Increased short-range connectivity in frontal cortex and atypicalities in resting-state EEG & Local over-connectivity \\\hline

        Predictive coding & Pellicano \& Burr (2012) \cite{pellicano2012world} & Excessive attention to detail & -- & Hypo-prior: lower precision of prior, stronger focus on sensory input \\\hline
        
        & Lawson et al. (2014) \cite{lawson2014aberrant} & Excessive attention to detail & Stronger activation in visual cortex than in prefrontal cortex in ASD & Hypo-prior or hyper sensory input: Precision imbalance that leads to excessive reliance on input \\\hline
        
        Recurrent NNs & H. Idei et al. (2017) \cite{idei2017reduced} & Stereotypical behaviors & -- & Modification of variance estimation (sensory precision)\\\hline
        
        & Philippsen \& Nagai (2018) \cite{philippsen2018understanding} & Reduced generalization capability, heterogeneity among subjects & -- & Modification of reliance on external signal and of variance estimation (sensory precision) \\\hline
        
        & Ahmadi \& Tani (2017) \cite{ahmadi2017bridging} & Generalization deficits & -- & Regularization \\\hline

        Other approaches & O'Loughlin and Thagard (2000) \cite{o2000autism} & Weak coherence, Theory of Mind impairment & -- & Impairment of coherence optimization in logical reasoning due to strong inhibition \\\hline

        \end{tabular}
        }
\label{table:autism}
\end{table*}

\subsection{Feed-forward and simple recurrent neural networks}

First, we describe approaches using simple connectionist models, typically feed-forward networks for classification tasks. Recurrent connections might be included at a structural level, but networks are not supposed to learn temporal sequences, which is why we refer to them as simple recurrent NN. These approaches mainly explored parameters of the network such as number of neurons or learning rate.

\subsubsection{Generalization deficits through overfitting}

\label{AutismCohen}

The first neural network model of ASD to our knowledge was proposed by Ira L. Cohen in 1994 \cite{cohen1994artificial}. It was a feed-forward neural network trained with back-propagation and investigated basic properties of neural networks. Based on studies that suggested that individuals with autism have either too few or too many neurons and neuronal connections (e.g., \cite{bauman1991microscopic}), the influence of increased or reduced number of hidden neurons was analyzed. The evaluated task was to classify children with ASD and children with mental retardation into two groups, using features obtained via a diagnostic interview \cite{cohen1993neural}. Note that although the considered task was related to ASD, the chosen task is just taken as an example and is not crucial for the findings of this paper.

A training and a test set were used to analyze the network's accuracy and generalization abilities. The results were compared for an increasing number of hidden units and through different number of trials. The results showed that a small number of hidden neurons translates into low accuracy (high training error) and bad generalization (high testing error) and an increased number of hidden neurons improved the network's learning accuracy and generalization. When the \textbf{number of hidden neurons was largely increased}, its generalization ability decreased: the network learned too much details of the input data and was not able to adapt to new input data.
An \textbf{increased number of training trials (longer training duration)} had a similar effect. For the training set, the network accuracy increased with longer training duration. However, with the test set, the network again showed signs of overfitting, as the accuracy decreased significantly.

Cohen compared these results qualitatively to the learning and behavioral characteristics of children with ASD.
In particular, many individuals with ASD show great discrimination capabilities and have no problems with already learned routines, but have problems when trying to abstract information or when confronted with new situations.

Cohen extended this approach in 1998 \cite{cohen1998neural} to the generalization capability in the presence of extraneous inputs to the network (set to random values). In the task of classifying happy and sad expressions of a simplified cartoon face, generalization was strongly impaired in the presence of extraneous inputs. This might suggest that networks trained for too long tend to attend more to non-relevant input information, instead of focusing on the more informative input neurons.

\annotation{Note that although increased number of hidden neurons may replicate autistic traits as shown in \cite{cohen1994artificial}, this parameter did not cause generalization deficits neither in Cohen's follow-up work \cite{cohen1998neural} nor in a similar modeling study \cite{dovgopoly2013connectionist} (see discussion on p.~\pageref{dovgopolyOverfitting}).}

\subsubsection{Precision of memory representations}

\label{AutismMcClelland}
In \cite{mcclelland2000basis}, James L. McClelland addressed the tendency of children with ASD to represent concepts in a too specific way, which results in difficulties to recognize two different instances of an object as the same category.


He suggested that in neural networks, this could be explained with the concept of excessive conjunctive coding. Typically, similar inputs to a neural network lead to similar neuron activation patterns. Such pattern overlaps can be useful for sharing existing knowledge and establishing associations. However, too strong associations can also cause interference. Conjunctive coding describes the reduction of such overlap by recoding the input patterns with neurons which only become active for particular combinations of elements. Assuming that what characterizes healthy human learning is a balance between generalization and discrimination, the representation of concepts in subjects with ASD could be characterized by \textbf{excessive conjunctive coding}. This would make a neural network loose the ability to generalize, as activation pattern overlaps cannot be exploited.

\begin{figure}
    \centering
        \includegraphics[width=0.9\columnwidth]{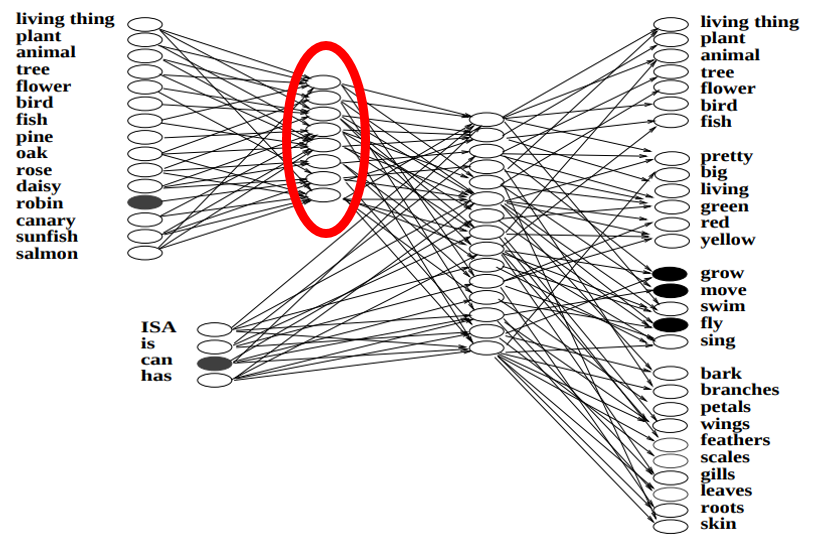}
    \caption{Semantic network used to explain the conjunctive coding hypothesis. In the hidden layers, the feed-forward neural network generates internal representations of the inputs (highlighted in red). Words describing similar concepts should produce similar internal representations that overlap with each other. The author suggests that excessive conjunctive coding to avoid these overlaps could produce excessive discrimination, such as in autistic perception. Adapted from \cite{mcclelland2000basis}.}
    \label{fig:McClellandFeedForward}
\end{figure}

This idea was not tested experimentally, but the author used the neural network shown in \fig{McClellandFeedForward} to explain his reasoning.
McClelland presented the example of a semantic network used in \cite{mcclelland1995there}, as a model of organization of knowledge in memory (see \fig{McClellandFeedForward}). This model was used to associate words with their meaning, e.g., ``robin" and ``can" trigger the outputs ``grow", ``move" and ``fly" because these are the actions a ``robin" can perform. The internal layer of the network (highlighted in red in \fig{McClellandFeedForward}) progressively learns to code the meaning of input words during learning. This means that ``robin" and ``canary" should cause a very similar activation pattern because a robin has much more in common with a canary than, for instance, a tree. The author suggests that hyperspecificity in perception and memory representations of ASD children might be caused by an abnormality during this process. Namely, excessive conjunctive coding in the internal layer is proposed as a mechanism: an excessive reduction of overlap between representations of similar concept might cause the reported hyperspecificity which would result in generalization deficits.
No concrete network parameters are proposed, but it can be imagined that such an effect might be achieved by increasing the number of neurons in the internal layer. In this regard, the approach is similar to Cohen's suggestion \cite{cohen1994artificial}, but extended to learning of representations.

\subsubsection{Generalization and categorization abilities in visual perception}

Dovgopoly and Mercado \cite{dovgopoly2013connectionist} used an existing model of visual object perception \cite{henderson2011pdp} to replicate deficits in classification and generalization in ASD. The neural network was a feed-forward network, which  modeled visual input processing via two pathways: the ventral cortical pathway (for object identification, including recurrent connections), and the dorsal cortical pathway (for processing of location-relevant information).

The authors replicated behavioral data from \cite{church2010atypical} and \cite{vladusich2010prototypical}, separately on both visual pathways, which show deficits in generalization and prototoype formation in children with high-functioning ASD. The experiment was the classification of random dot patterns as category or non-category stimuli \cite{church2010atypical}, or as category A or category B stimuli \cite{vladusich2010prototypical}. After adjusting the parameters for replicating typical behavior, four different parameter modifications were tested individually to replicate the data from ASD children. Following evidence for abnormalities in synaptic plasticity in individuals with ASD (e.g., \cite{bourgeron2009synaptic, auerbach2011mutations}), the first two parameters modified how weights in the network were updated.

First, the \textbf{learning rate was decreased}, which corresponds to reduced synaptic plasticity in biological neurons. As a result, network training takes longer and is more prone to lead to exhibit overfitting. Second, \textbf{generalization of the network was impaired by suppressing regularization} using negative weight decay. Weight decay is a method for regularizing neural networks and improving their generalization abilities by keeping the connection weights small \cite{krogh1992simple}. Typically, weight decay punishes large weights by adding a term $\lambda \vec{w}'\vec{w}$ to the error function. With a negative weight decay factor $\lambda$ instead, anti-regularization is performed, encouraging the increase of weight magnitudes, and thus, over-complex classification rules. Third, they tested the influence of \textbf{increasing and decreasing the number of hidden neurons} similar to \cite{cohen1994artificial, cohen1998neural}, based on neurological evidence of an increased number of cortical minicolumns in the brain of individuals with ASD \cite{casanova2006minicolumnar}. Finally, the authors adjusted the \textbf{gain of the neuron's activation function, to model the increased level of noise} that is hypothesized to underlie the relative increase in cortical excitation observed in ASD subjects \cite{rubenstein2003model, yizhar2011neocortical}.


The gain $G$ of the activation function, as displayed in \eq{gain}, manipulates the slope of the activation function. A smaller gain reduces the slope, and makes the network more prone to pass noise instead of signal information to the next processing layers:

\begin{equation}
    s(x) = \frac{1}{1 + \exp(-(G \cdot x + b))}
    \label{eq:gain}
\end{equation}
where $x$ represents the input to the activation function and $b$ is a bias term.


\label{dovgopolyOverfitting}
Good replications of the behavioral data were achieved with a decrease of learning rate and a negative weight decay. A negative weight decay also caused a high variability of generalization abilities, depending on the initial network weights, providing a potential explanation for the heterogeneity of findings between different studies.
The gain of the activation function could not fully account for the generalization deficit.
Also an increased number of neurons did not replicate the generalization deficit in ASD children, which contradicts previous findings from \cite{cohen1994artificial}.
\annotation{In fact, an increased number of hidden units seems to lead to generalization problems only under certain training circumstances \cite{caruana2001overfitting}, indicating that it is not a good candidate for explaining generalization difficulties in general.}



\subsection{Self-organizing maps}
Self-organizing maps (SOMs) are ANNs that are usually used for unsupervised learning and clustering tasks. They model the functionality of cortical feature maps, which are spatially organized neurons that respond to stimuli and self-organize according to the features in stimuli. They are able to learn the relation of different input data such as different sensory inputs. Approaches for modeling ASD with SOMs typically investigate the formation of higher-level representations from sensory input.



\subsubsection{Increased lateral feedback inhibition}
\label{AutismGustafsson1}

Lennart Gustafsson presented two models of ASD using SOMs in \cite{gustafsson1997inadequate} and \cite{gustafsson2004self}. Inspired by findings on weak central coherence in subjects with ASD and an enhanced ability to discriminate sensory stimuli \cite{frith1994autism}, he suggested that alterations in the lateral feedback weights between the SOM neurons could result in atypicalities in perception \cite{mountcastle1957modality}.

\begin{figure}[t]
    \centering
        \includegraphics[width=\columnwidth]{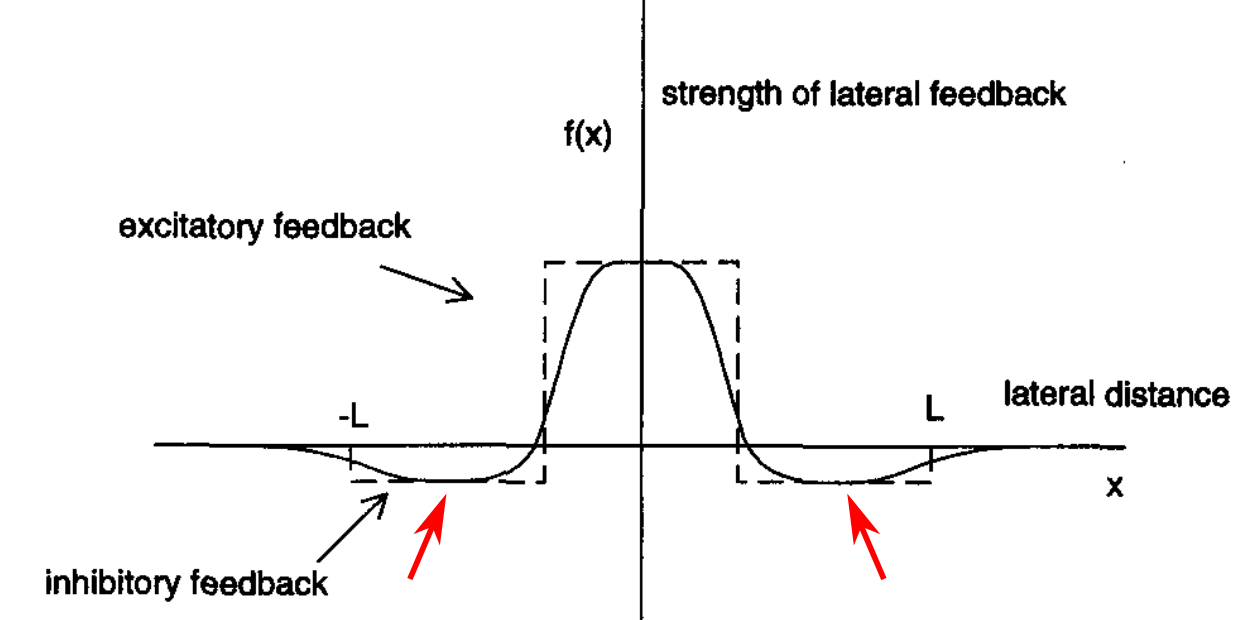}
    \caption{Mexican-hat function of the SOM. It defines the strength of lateral connections depending on distance to current neuron. The red arrows point to the part that is modified to simulate autistic perception (excessive lateral feedback inhibition). Adapted from \cite{gustafsson1997inadequate}.}
    \label{fig:Gustafsson1997}
\end{figure}

In a SOM, each neuron typically has excitatory connections to close neighbors and inhibitory connections to more distant neighbors. They tuned the Mexican-hat curve (\fig{Gustafsson1997}) to induce stronger lateral feedback inhibition. Such activation patterns are similar to receptive fields in biological cortices and have been used to model center-surround operators in the visual cortex. Manipulating the lateral connections to achieve a stronger inhibition (such that the integral of the function in \fig{Gustafsson1997} becomes negative), the sensory discrimination ability of the network is increased. Neural columns focus on more narrow features during learning which slows down convergence and might lead to a fragmented feature map. However, excessive lateral inhibition will degrade discriminatory power and cause instabilities in information processing. This behavior is compared to autistic over-discrimination and may also explain fascination or fright of moving objects, due to the instability of its cortical feature maps.

\subsubsection{Familiarity preference}

\label{AutismGustafsson2}


In \cite{gustafsson2004self}, Gustafsson and Papli{\'n}ski evaluated the effect of attention-shift impairment and avoidance of novelty on the formation of cortical feature maps. The used SOM received input stimuli from two sources (compared to two ``dialects of a language''), each of which produces 30 different stimuli (``speech sounds'') grouped in three clusters (``phonemes'').

The computational model was run in four different modes. In the first mode, attention was always shifted to the source producing novel input (considered as normal learning). In the second mode, an attention-shift impairment was modeled by shifting attention to novel sources with a very low probability. The third mode implements familiarity preference: attention is shifted to novel sources only if the map is familiar with that source (measured as mean distance of the current stimulus to the map nodes). This map develops a preference over learning to the more familiar source. Finally, a model with both familiarity preference and attention-shift impairment was applied.

The simulation results showed that \textbf{familiarity preference} leads to precise learning of the stimuli from one of the sources (the source with lower variability) in expense of the other source. This might remind of ASD individuals' characteristic of learning in great detail a narrow field, which leads to increased discrimination and poor generalization. The authors also showed that this impairment can be counteracted by modifying the probabilities of stimuli presentation in response to the system, similar to early intervention in children's learning process. Maps learned with attention-shift were not impaired, whereas a combination of both mechanisms only sometimes led to an impairment. The authors concluded that, in contrast to speculations in previous work \cite{courchesne1994impairment}, familiarity preference, rather than attention-shift is a more likely cause for ASD.

\subsubsection{Unfolding of feature maps and stimuli coverage}

In 2007, Gerardo Noriega \cite{noriega2007self} modeled abnormalities in the feature coverage and the unfolding of feature maps in SOMs. Neurological evidence suggests abnormal brain development in children with ASD \cite{bauman2005neuroanatomic}, typically reporting larger growth in young children, which gets reduced later in life \cite{courchesne2001unusual, aylward2002effects}. These abnormalities were modeled by manipulating the number of network nodes during the training of the SOM where the structure emerges. Thus, the network dimension is temporarily increased.

Results showed that such disturbance in the physical structure of a SOM does not affect stimuli coverage, but impairs the unfolding of feature maps which might result in sub-optimal representations. Furthermore, the author models hyper- and hyposensitivity to stimuli in a similar way like \cite{gustafsson1997inadequate} using lateral interactions between neurons. \textbf{Hyper- or hyposensitivity} was modeled by adjusting the neuron weights toward the winner neuron, either with a positive factor (attraction, or hypersensitivity) or with a negative factor (repulsion, or hyposensitivity). This factor converges exponentially toward zero (normal sensitivity) during map formation. The authors showed that hypersensitivity to one of the input domains (stronger attention to this domain, i.e., restricted interests), improves the coverage of stimuli in this domain, but too strong hypersensitivity or a hyposensitivity to stimuli reduces coverage \footnote{Hypersensitivity in \cite{gustafsson1997inadequate} was implemented as increased inhibition in the neighborhood of neurons (higher specificity of perception), whereas this approach interprets hypersensitivity as a stronger attraction of neighboring signals to signals from a specific domain.}.



One year later, Noriega extended his approach in \cite{noriega2008modeling}, investigating \textbf{propagation delays between neurons}.
Unlike in normal SOMs where all neurons propagate the information instantaneously to all neighboring neurons, Noriega presented a biologically more realistic approach by introducing delays in the update.
He shows that decreased propagation speed has a negative effect on stimuli coverage.
As the delayed propagation causes the arrival of competing stimuli at the same time at a neuron, he also altered the way in which these competing stimuli are handled. In his experiments, a high \textit{dilution factor}, meaning that incoming stimuli are averaged instead of being handled separately, decreased the stimuli coverage and also impaired the topological structure of the map.

%

\subsection{Convolutional neural networks and inhibition imbalance}


In 2015, Y. Nagai and colleagues presented an ANN network based on Fukushima's neocognitron (\cite{fukushima1982neocognitron}, \cite{fukushima1988neocognitron}, \cite{fukushima2003neocognitron}), seen as the basis for convolutional neural networks, to model visual processing in ASD \cite{nagai2015influence}. The hypothesis considered was that there is an excitation/inhibition imbalance in ASD \cite{sun2012impaired,snijders2013atypical,yizhar2011neocortical}.

\begin{figure}[hbtp!]
    \centering
        \includegraphics[width=\textwidth]{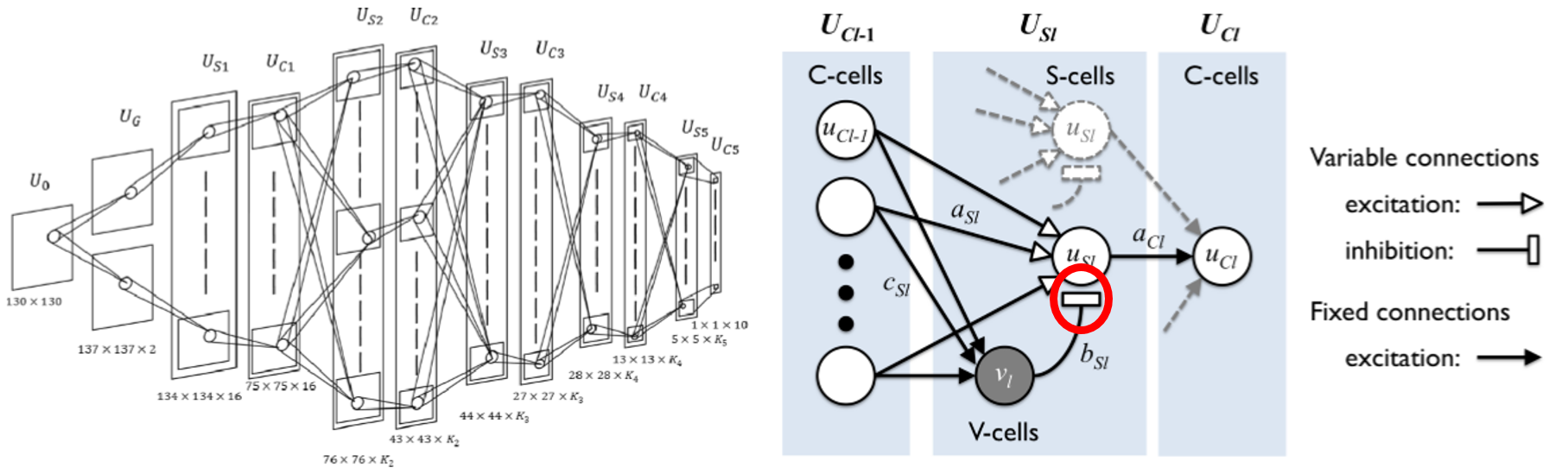}
    \caption{Left: Overview of the neocognitron's structure. Right: Detailed view of the connections between C-cell layers $U_C$ and S-cell layers $U_S$. Highlighted in red are the inhibitory connections that are modified to influence the ratio between inhibition and excitation. Adapted from \cite{nagai2015influence}.}
    \label{fig:NagaiAutism}
\end{figure}

The structure of the neocognitron for visual processing is illustrated in \fig{NagaiAutism}.
The network is trained to recognize patterns by adjusting the weights between $U_C$ and $U_S$ layers.
The S-cells in the $U_S$ layers perform feature extraction. They receive excitatory input from the C-cells in the preceding layer, and inhibitory connections from the V-cells in the same layer. During training, the excitatory connections $a_{Sl}$ are updated and the inhibitory connections $b_{Sl}$ are calculated accordingly.

\begin{figure}[hbtp!]
    \centering
        \includegraphics[scale=0.65]{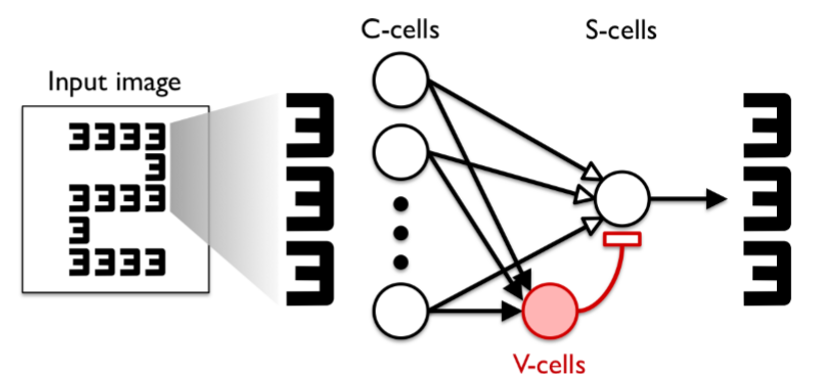}
    \caption{The neocognitron is fed with a visual stimulus consisting of local patterns (here \textit{3}) and global patterns (here \textit{2}), which are incongruent. In normal conditions the network should be able to detect both local and global patterns.} 
    \label{fig:NagaiAutism2}
\end{figure}

The network was trained for the recognition of numbers ``0'' to ``9'' in large or small size at different positions.
After training, the model was tested with compound numbers (cf. \fig{NagaiAutism2} left) where a larger number is created from multiple smaller numbers.
The trained network is able to detect both global (large number, here ``2'') and local (small numbers, here ``3'') patterns for $\alpha = 1$ and $0.9$, but shows a preference for the global pattern, characteristics that correspond to observations with healthy individuals \cite{behrmann2006configural}.


It is known that people with ASD perform differently in such a task, primarily focusing their attention on the details (i.e., the smaller number instead of the larger one).
In order to simulate this local processing bias, an imbalance of excitatory and inhibitory connections was simulated by scaling the \textbf{inhibitory weight $\bm{b_{Sl}}$ with a factor $\bm{\alpha}$}.


The results show that a moderate increase of $\alpha$, which corresponds to increasing inhibition, causes the network to rather detect local patterns, replicating the local processing bias in ASD.
When reducing $\alpha$ (increasing excitation), the network does not show any processing bias, rather it looses its ability to differentiate patterns.
These results fit with ASD symptoms of hyperesthesia (increased focus on detail) and hypoesthesia (no bias and general difficulty in pattern recognition) and suggest that excitation/inhibition imbalance could account for these symptoms.

\subsection{Spiking neural networks and local over-connectivity}


In \cite{ichinose2017local} and a follow-up study in \cite{park2019macroscopic}, it was proposed to use spiking neural network as computational models to investigate the consequences of local over-connectivity, which was found in the prefrontal cortex of ASD brains \cite{courchesne2005frontal}. The hypothesis considered was that local over-connectivity affects frequency patterns of neural activations.

A spiking neural network is more closely inspired by natural neural networks \cite{izhikevich2003simple}. Whereas in standard artificial neural networks each neuron fires at every time step, neurons in a spiking network only fire if their potential (similar to the membrane potential of biological neurons) exceeds a certain threshold. Therefore, more complex firing patterns can occur ranging over various frequency bands, comparable to patterns visible in EEG\footnote{EEG: Electroencephalography}.

\annotation{A number of studies found evidence that EEG signals of ASD brains tend to exhibit higher power in low-frequency and high-frequency bands of EEG \cite{wang2013resting} and that EEG resting-state activity has lower complexity \cite{bosl2011eeg}.
The authors suggest that these atypical EEG data might be explained by differences in how ASD brains, as opposed to TD brains, are connected. In particular, it has been found that the brains of people with ASD have an increased local connectivity, especially in the frontal cortex \cite{courchesne2005frontal}.}

\begin{figure}
    \centering
    \includegraphics[width=\columnwidth]{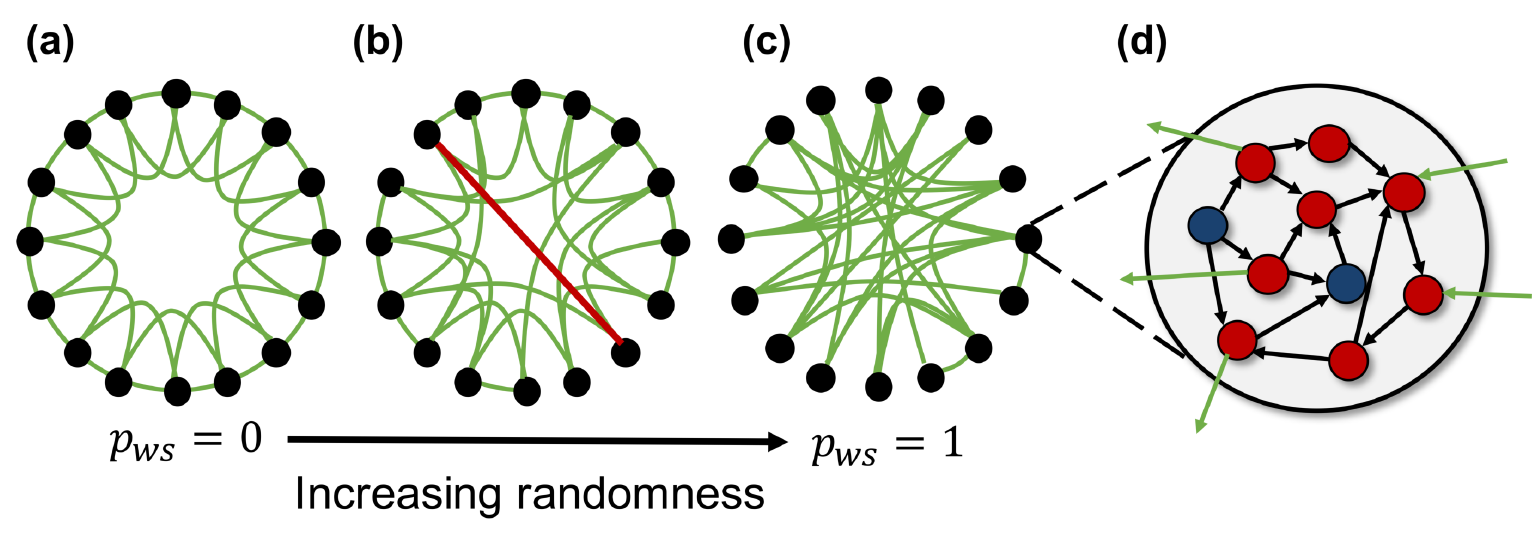}
    \caption{Three different networks  with different degrees of randomness. (a) is a locally over-connected network (corresponding to ASD individuals), (b) is a small-world network with many local clusters and a few longer connections (corresponding to typically developed individuals), (c) is a random network including many wide-range connections. (d) shows the structure of each single neuron group with excitatory (red) and inhibitory (blue) connections. Note that the number of nodes and edges in (a), (b) and (c) remains the same. Reprinted with permission from \cite{park2019macroscopic}, originally based on \cite{watts1998collective}.}
    \label{fig:AsadaAutism}
\end{figure}

The authors investigated this hypothesis with a spiking neural network by modifying the network's connection patterns and observing how the connectivity affected the emerged activation patterns.
To manipulate the \textbf{degree of local over-connectivity in the network}, a parameter based on the small-world paradigm from \cite{watts1998collective} was used. By default, neurons are connected to six neighboring neurons in a ring lattice as displayed in \fig{AsadaAutism} (left). A parameter $p_{WS}$ expresses the probability for each of the connections to rewire to other neurons. Thus, $p_{WS}$ determines the randomness of the network (\fig{AsadaAutism}), ranging from regular lattice structure ($p_{WS} = 0$) to random wiring ($p_{WS} = 1$). Medium values of $p_{WS}$ around $0.2$ describe ``typically developed networks'' with local clusters and some short-range connections between the clusters.
Notably, the parameter from \cite{watts1998collective} keeps the overall number of connections in the network intact, such that differences emerge only due to differences in the network structure, not by the total number of neurons or neural connections.

Networks are formed by generating 100 groups of neurons, corresponding to the black nodes in \fig{AsadaAutism}. Each group contains 1000 spiking neurons: 800 excitatory and 200 inhibitory neurons, which have an increasing or decreasing effect on the firing probability of postsynaptic neurons, respectively. Neurons are mainly connected to neurons of the same neuron group (intra-group connections), and have connections to six neighboring groups according to \fig{AsadaAutism} (inter-group connections). Different rewiring probabilities $p_{WS}$ between $0$ and $1$ are used to determine the initial inter-group connectivity of the network.

After initialization, the network updates its connections according to the rules of spike-time-dependent plasticity \cite{izhikevich2003relating}: the update of connection weights occurs depending on the timing of firing of the pre- and postsynaptic neurons. If the postsynaptic neuron fires within a certain time window \textit{after the presynaptic neuron}, the weight of the connection is increased (corresponding to the biological process of \textit{long term potentiation}). If the presynaptic neuron fires within a time window \textit{after the postsynaptic neuron}, the connection weight is weakened (\textit{long term depression}).
During this learning period the connection weights self-organize. Tonic random input is presented to the network. After learning, the spontaneous activity of the neurons was recorded (in the absence of input), and compared to the graph-theoretical properties of the network.

The activation patterns were evaluated according to their frequency spectrum and the complexity of the time series, as measured by the multiscale entropy \cite{costa2005multiscale}. This measure rates the informative content of time series at different temporal scales. High complexity corresponds to the presence of long-range correlations on multiple scales in space and time, low complexity is computed for time-series with perfect regularity or randomness. The evaluation suggested that networks exhibiting local over-connectivity generate more oscillations in high-frequency bands and exhibit lower complexity in the signals than small-world networks. Findings of atypical resting-state EEG for people with ASD, thus, might be explained by local over-connectivity in their brains.


\subsection{Bayesian approaches}
\label{sec:bayesianASD}

There are promising models in the literature interpreting ASD on the basis of the Bayesian framework (for an introduction see Schizophrenia section, p.~\pageref{sec:bayesianSchizo}). However, most of these approaches are only conceptual and still lack an implementation. Nevertheless, these approaches are able to explain a wide range of different symptoms which might be caused by an atypical integration of prediction and sensory information \cite{lawson2014aberrant, pellicano2012world}.



The first approach utilizing the Bayesian brain hypothesis for explaining the non-social symptoms of ASD was proposed by Pellicano and Burr in 2012 \cite{pellicano2012world}.
Their \textbf{hypo-prior hypothesis}\footnote{In this article, we stick to the original definition of hypo-priors as a belief in low precision of priors and hyper-priors as a belief in high precision of priors. Note, however, that due to the hierarchical structure of the brain and the role of precision as a hyperparameter for the inference process it might be more appropriate to talk of hypo-priors as attenuated hyperpriors as argued in \cite{friston2013hyperpriors}.} suggests that broader or less precise priors cause people with ASD to rely less on their predictions and stronger on sensory input which could explain the hypersensitivity of people with ASD.
J. Brock broadened this idea \cite{brock2012alternative} by proposing that hypersensitivity cannot only be caused by a reduced precision of the prior, but also by an increased precision of sensory input.
Lawson et al. \cite{lawson2014aberrant} summarized these ideas, arguing that both modifications \textbf{reduced prior precision or increased sensory precision}, can cause the same functional consequences.
They suggest that the cause could be aberrant precision in general:
Expected precision of a signal is an important source of information that helps us to decide whether to rely on this signal or not. Aberrant precision of sensory input or prior predictions, thus, would alter the way in which we integrate these signals.
The precision of the signals also can be considered as a weighting term of the prediction error: For a signal that is expected to be imprecise, a prediction error does not need to be corrected while a prediction error arising between signals that are expected to be very precise would need correction.
People with ASD might have problems to accurately estimate this precision. Thus, they might, at the one extreme, try to minimize the prediction error too strongly, or, at the other extreme, fail to minimize the prediction error.

\annotation{Finally, in \cite{lawson2017adults}, Lawson and colleagues suggested that subjects with ASD overestimate the volatility of the environment.
They conducted a behavioral experiment which demonstrated that ASD subjects are less surprised when encountering environmental changes.
Using Hierarchical Gaussian Filters, they modeled the experimental findings computationally. The model parameter that best accounts for the differences found in ASD and neurotypical subjects was a meta-parameter which controlled learning about volatility of the environment.
These results suggest that ASD subjects overestimate the probability of a change in the environmental conditions, and build less stable expectations. As a result, they might misinterpret an event with low probability which occurred by chance as an event that signifies a change in environmental conditions. Therefore, instead of being surprised in the case of an extraordinary event, they would be mildly surprised at all times.}





\subsection{Recurrent neural networks}

The studies presented here follow the idea of predictive coding which can be seen as an implementation of the Bayesian brain idea: an RNN is used as an internal model of the world and its learning corresponds to the process of adapting network weights in order to perform prediction error minimization.
The role of the network is to learn to predict sensory consequences, and integrates these predictions with the perceived sensory information.


\subsubsection{Freezing and repetitive behavior in a robotics experiment}
\label{AutismTani}

\begin{figure*}
    \centering
        \includegraphics[width=\textwidth]{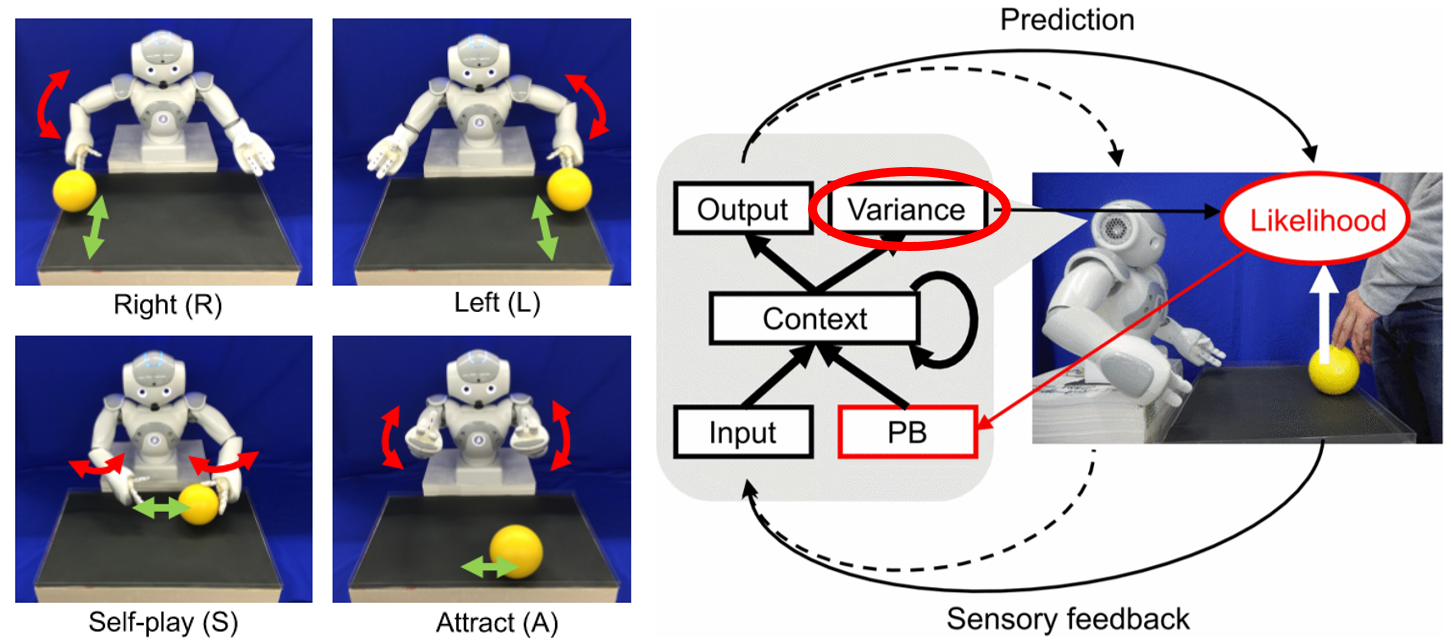}
    \caption{Left: Overview of the interactive tasks the robot must perform. Right: Overview of the ANN model used for the experiments. Highlighted in red are the variance units where a constant $K$ is added to increase or decrease the sensory precision in order to imitate autistic behavior. Adapted from \cite{idei2017reduced}.}
    \label{fig:TaniAutism}
\end{figure*}

Idei and colleagues \cite{idei2017reduced, idei2018neurorobotics} used the stochastic continuous-time recurrent neural network (S-CTRNN) \cite{murata2013learning} model with parametric bias (PB) \cite{tani2003learning} to teach a robot to interact with a human in a ball-playing game (similar to the schizophrenia model \cite{yamashita2012spontaneous}).
The S-CTRNN with PB learns to predict a time series of proprioceptive (joint angles) and vision features. From the current input, the network estimates the next time step (output) and its predicted precision (variance) as  shown in \fig{TaniAutism}.
The state of the PB units reflect the intention of the network, i.e., the ball-playing pattern that the robot believes that they are currently engaged in.

The S-CTRNN was trained offline to perform certain tasks depending on a yellow ball's position, as depicted in \fig{TaniAutism} (left). Synaptic weights and biases of the network, as well as the internal states of the PB units are updated via the backpropagation through time (BPTT) algorithm in order to maximize the likelihood in \eq{LikelihoodFunction}.
This equation describes that  at time step $t$ of training sequence $s$, the network output of the $i$-th neuron (a normal distribution defined by the estimated mean (output) $y$ and estimated variance $v$) properly reflects the desired input data $\hat{y}$.

\begin{equation}
    L_{t,i}^{(s)} = -\frac{\ln{(2\pi v_{t,i}^{(s)}})}{2} - \frac{(\hat{y}_{t,i}^{(s)}-y_{t,i}^{(s)})^2}{2v_{t,i}^{(s)}}
    \label{eq:LikelihoodFunction}
\end{equation}
After training, a recognition mechanism (via adaptation of the PB units, while keeping weights and biases fixed) enables the network to switch its behavior depending on the current situation.







To model ASD behavior, the \textbf{estimated variance (sensory precision) is modified} in the activation function of the variance units with the constant $K$ in \eq{VarianceModification}, where $\epsilon$ is the minimum value and $u_{t,i}^{(s)}$ is the output of the $i$-th context unit time step $t$ for movement sequence $s$.  

\begin{equation}
    v_{t,i}^{(s)} = \exp(u_{t,i}^{(s)} + K) + \epsilon
    \label{eq:VarianceModification}
\end{equation}

Experimental results with a humanoid NAO robot showed that for $K = 0$ the robot behaved normally. For increased variance (reduced precision), the robot seemed to ignore prediction error and performed stopping and stereotypic movements. For decreased variance (increased precision), the robot performed incorrect movement changes or concentrated on certain movements, which also led to sudden freezing and repetitive movements. These results fit with the disordered motor system reported in ASD \cite{gowen2013motor}, but add the surprising insight that increased and decreased sensory precision may cause the same consequences.

\subsubsection{Impairment in internal network representations}

Another study using the S-CTRNN to model ASD characteristics is \cite{philippsen2018understanding}.
Using an S-CTRNN \cite{murata2013learning}, the authors modify two parameters which control how the network makes predictions.
In contrast to the other RNN model which concentrates on replicating behavioral patterns, this study investigates ``invisible'' features characterizing the network's learning process.
More specifically, the authors evaluate \textbf{how attention to sensory input and deficits in the prediction of trajectory noise influence the internal representation} that a network acquires during learning.
\annotation{Internal representations are informative as they reflect the network's generalization capabilities \cite{boden2002guide, yamashita2008emergence}: similar input pattern should cause an overlap in the corresponding context neuron activations (attractors in the RNN), whereas different patterns should be differentiated.}

\begin{figure}
    \centering
    \includegraphics[width=\columnwidth]{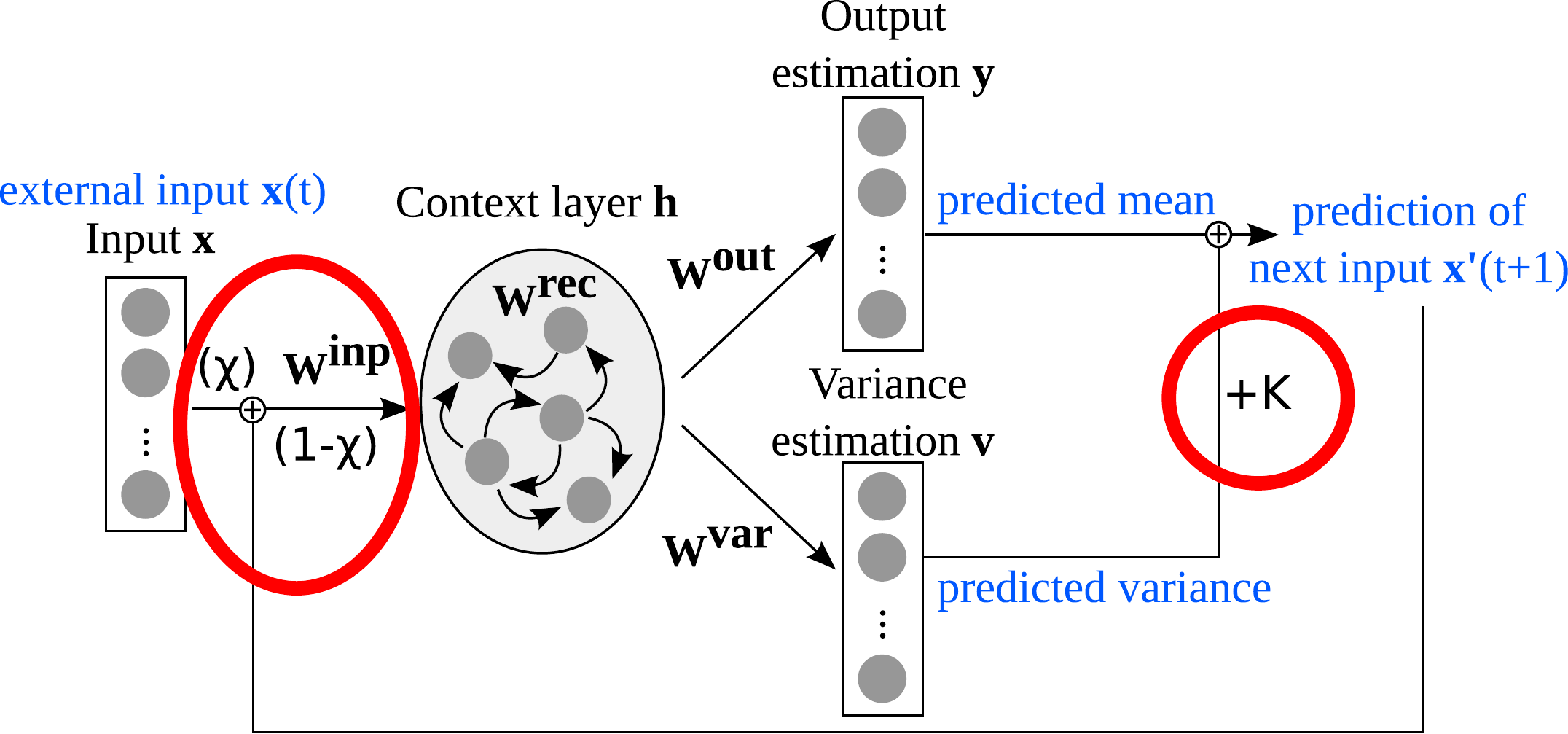}
    \caption{The S-CTRNN used in \cite{philippsen2018understanding} with two parameter modifications. Adapted from \cite{philippsen2018understanding}.}
    \label{fig:philippsenAutism}
\end{figure}

\begin{figure}[hbpt!]
    \centering
    \subfigure[Hypothesis]{
    \includegraphics[width=0.7\linewidth]{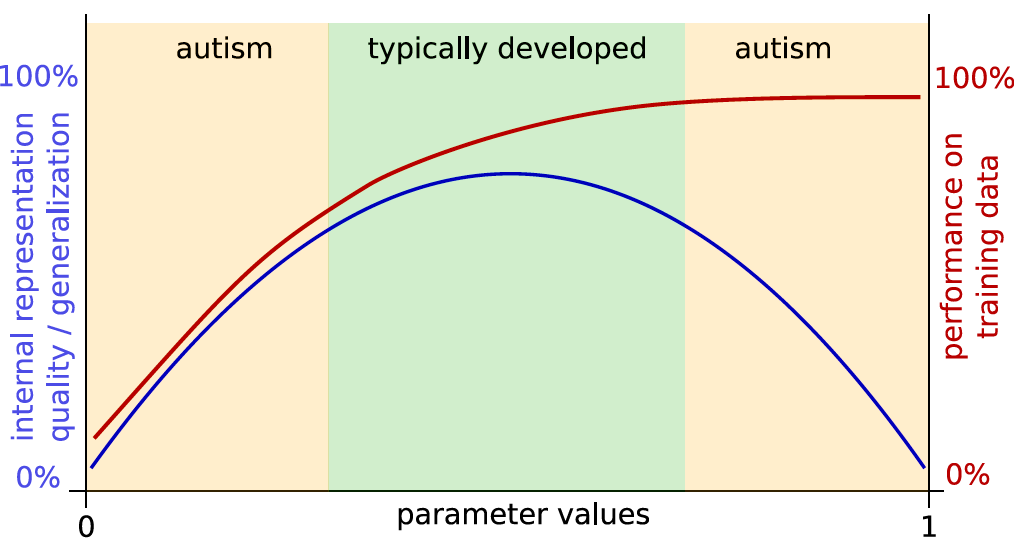}
    \label{fig:philippsenAutism2:hypo}
    }\\
    \subfigure[Experimental results]{
    \includegraphics[width=0.7\linewidth]{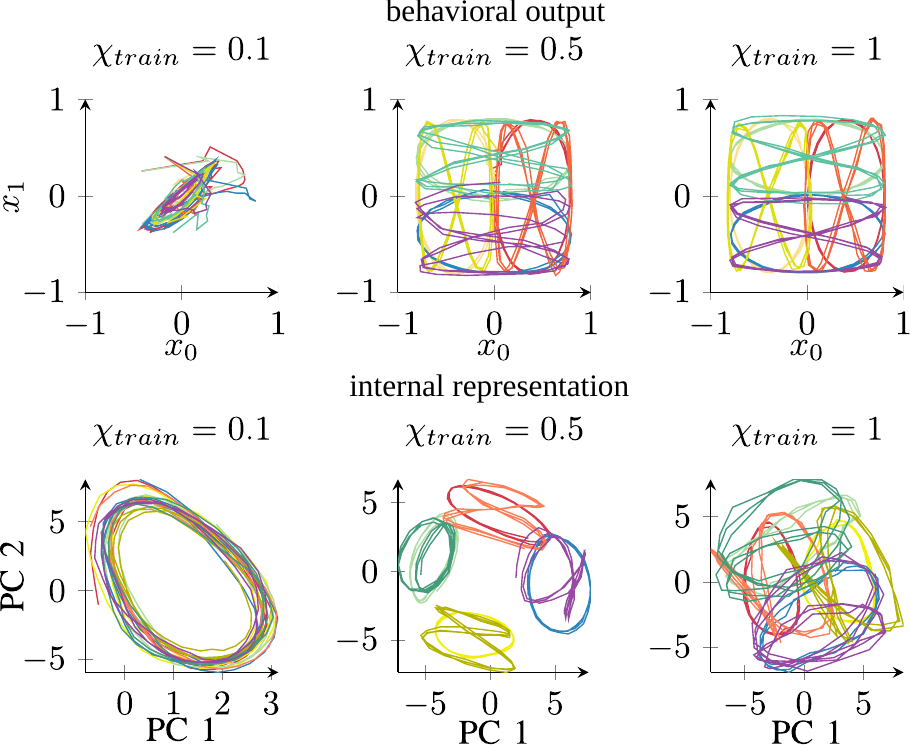}
    \label{fig:philippsenAutism2:chi}
    }
    
    \caption{Effect of changing the external contribution parameter of the S-CTRNN from \fig{philippsenAutism} on behavioral output (top) and on internal representation quality, evaluated in the two-dimenisonal principal components (PC) space (bottom). Adapted from \cite{philippsen2018understanding}.}
    \label{fig:philippsenAutism2}
\end{figure}

The network as displayed in \fig{philippsenAutism} is trained to recognize and draw ellipses and ``eight'' shapes, located at four different (overlapping) positions of the input space (cf. \fig{philippsenAutism2:chi}).
Inputs and outputs are two-dimensional trajectories and the recurrent context layer comprises $70$ neurons.
Learning is modified in two ways: The parameter $\chi$ determines how much the network relies on external input, as opposed to its own prediction, i.e., $\chi$ gradually switches between open-loop ($\chi = 1$) and near-closed-loop  ($\chi \approx 0$) control.
The second parameter $K$ is defined analogous to \cite{idei2017reduced} (see \eq{VarianceModification}) and manipulates the estimated variance such that networks with $K \neq 0$ over- or underestimate noisy variations in the signal. Unlike its usage in \cite{idei2017reduced}, this manipulation is not performed after training, but already \textit{during} the training process, to account for the developmental nature of ASD.

After training, the network's behavior is evaluated as the network's ability to reproduce the trained trajectories.
The internal representations are evaluated by collecting the time course of activations of the context layer neurons while generating the trajectories.

A visualization of how the high-dimensional space (time steps $\times$ number of context neurons) is structured can be achieved by principal component analysis (PCA). The results indicated that networks tend to reuse internal representation structure for patterns located at the same position in the input space.
\annotation{Such an overlap is advantageous as similarities between patterns are coded. However, too strong overlap of the context activations indicates missing differentiation between the patterns which might lead to worse differentiation in a recognition task.}
Thus, the authors define ``good'' internal network representations as representations which strongly reflect the characteristics of the input data.
\fig{philippsenAutism2:chi} shows an example of how task performance (top) and internal representation quality (bottom) change depending on the external contribution parameter.
The best internal representation quality is achieved with $\chi = 0.5$ \annotation{(moderately integrating input and predictions), as the internal representation reuses activations but clearly differentiates trajectories at different input space positions.} 
\annotation{However, the performance in reproducing the trained behavior is comparable between $\chi = 0.5$ and $\chi = 1$ (relying stronger on input).}
These qualitative observations were also quantitatively verified in the high-dimensional space of neurons.
\annotation{How well the network is able to reproduce the learned patterns, thus, is not always reflected in the internal representation quality.}


Interestingly, for the parameter $\chi$, both extremes lead to an ASD-like impairment, as schematically depicted in \fig{philippsenAutism2:hypo}. Typical development could correspond to the middle. Whereas the right-hand side would express high-functioning ASD where \annotation{the performance in specific tasks might be intact, but representations might be too specific (overfitting)}. The left-hand side describes ASD with severe impairments also at a behavioral level.
\annotation{It can be, thus, imagined that heterogeneity in the ASD population, comprising opposite symptoms such as hyper- and hyposensitivity, does not necessarily be caused by different underlying mechanisms, but that a continuous modification of parameters could account for the variability.}

\subsubsection{Generalization ability in a variational Bayes recurrent neural network}

In \cite{ahmadi2017bridging}, a novel recurrent network type is introduced, the variational Bayes predictive coding RNN (VBP-RNN).
It differs from the S-CTRNN in that variance is not only coded on the output level, but also in the network's context neurons to enhance the network's ability to represent uncertainty in the data.

We do not discuss it in detail here, as this study is not focusing on modeling ASD, but on representing deterministic as well as probabilistic behavior in an RNN in a coherent way. The analogy to ASD is made in terms of the \textbf{meta-parameter $W$ that performs a trade-off between reconstruction and regularization in the optimization (loss) function}. $W$ switches between the typically minimized reconstruction error term ($W=0$) and a regularization term that keeps the posterior distribution of the latent variables (i.e., the context units) similar to its prior.
If the network is trained with $W=0$, it develops deterministic dynamics and exhibits poor generalization capabilities. Values of $W>0$ lead to more randomness in the network and improve generalization, but too high values result in a performance drop.

$W$ could therefore model the spectrum of ASD: $W=0$ is one extreme where the network solely relies on its top-down intentionality and fails to generalize, whereas too high values of $W$ reflect performance impairment due to excessive randomness in the network.
As this parameter controls how much regularization is performed, the approach is similar to \cite{dovgopoly2013connectionist} where regularization was intentionally impaired.


\subsection{Other approaches}



In 2000, O'Loughlin and Thagard \cite{o2000autism} used a connectionist model to simulate weak coherence, and to demonstrate how a failure of maximizing global coherence can cause deficits in theory of mind \cite{baron1997mindblindness}. Their network model, a so-called constraint network, is hand-designed according to the task and does not strictly fit into an existing network category.
The network performs logical reasoning and consists of a set of neurons, each of which corresponds to a logical element such as a belief (expressed as a sentence).
Connections between them are set as excitatory and inhibitory, depending on whether two arguments support each other or are contradicting.
Weights remain fixed, but the activations of neurons get updated depending on the connections to neighboring cells which can be excitatory (positive) on inhibitory (negative).
A decaying factor lets the network's activation converge to a state after a certain amount of time. Positive activations are then interpreted as an acceptance of this belief, negative activations as a denial.

The authors showed that \textbf{a high level of inhibition, compared to excitation}, causes early activated association nodes in the network to suppress concurring hypotheses. The network, therefore, prefers more direct solutions, and makes wrong predictions. The overall \textit{coherence} of the network, defined as the satisfaction of most constraints in the complete network, is not optimized, which can be considered as weak coherence.

\section{Discussion and future directions}
\label{sec:new}

Artificial neural network models of SZ and ASD have been presented as a useful tool to fill the gap between theoretical models and biological evidence. Early works were biased by technical restrictions, but recent models are able to capture the same complexity as conceptual models, such as hierarchical Bayesian models. \annotation{However, designing ANN architectures that are able to predict novel findings and through computational simulations contribute to clinical applications (e.g., diagnosis or therapy) remains a challenging task.}
\annotation{In this sense, the model should i) reproduce empirical behavioural findings, preferably in more than one domain, ii) be supported by a process theory in which the abnormality used to reproduce empirical findings is realistic from the point of view of known neuropathology, and iii) predict novel findings. Furthermore, addressing heterogeneity and non-specificity is still one of the most important challenges of these two psychiatric disorders.}

\annotation{Due to the large overlap in SZ and ASD regarding biological evidence (e.g., E/I imbalance), similar hypotheses were discussed as a potential cause for both disorders.
Computational models, however, still tend to focus on specific impairments of a specific disorder.
To help the community, it is crucial that overarching neural network models are developed which connect ideas and results across different contexts (ASD, SZ or even other mental disorders).}


\annotation{In this section, we first discuss the quality of the discussed models in terms of how well they fit and predict empirical findings (\refsec{discussion:quality}).
Secondly, we discuss the approaches from the point of view of multifinality and equifinality (\refsec{discussion:multifinality}).
Thirdly, we emphasize the importance of testing the models in an embodied system (\refsec{discussion:realworld}).
Finally, we describe new promising directions to address with ANN models: developmental factors (\refsec{discussion:developmental}); disorders of the self (\refsec{discussion:self}); and state-of-the-art ANN architectures for future models of psychopathologies (\refsec{discussion:ANN}).}



\subsection{Models quality: Empirical findings and predictability}
\label{sec:discussion:quality}

\annotation{Early SZ modelling works from \cite{hoffman1989cortical, ruppin1995neural} on Hopfield networks as well as the feed-forward approaches from \cite{cohen1992context} and \cite{hoffman1997synaptic} lack the capabilities to generalize to a broader context: every experiment required a different ANN architecture. Hence, in terms of predictability of other symptoms, these approaches are not powerful enough. In particular, the work on auditory hallucinations \cite{hoffman1997synaptic} is far from replicating the brain mechanism and does not account for deficits in distinguishing self-produced sounds observed in SZ patients. However, the underlying discussion presented in those papers still provides valuable insights. They highlighted the connectivity factor between different cortical areas of the brain (either by gain reduction or pruning) specially in the context ones. Later works on RNN, such as \cite{yamashita2012spontaneous}, revisited this idea with hierarchical networks, with the same capability to generate parasitic states due to dynamic attractors. Pruning was substituted by noise injection. Interestingly, there are conceptual similarities between noise injection and precision reduction used in Bayesian approaches. Due to the more general architecture regarding sensorimotor integration, this RNN might be able to replicate other findings in earlier works such as hallucinations or performance in the Stroop task, however, this has not been experimentally demonstrated yet.}


\annotation{Bayesian approaches, such as predictive processing \cite{adams2013computational} and circular inference \cite{jardri2013circular} have shown better quality in terms of predictability of new empirical findings. Their mathematical abstraction is more powerful and may be applicable to different types of experiments. For instance, within the free-energy optimization framework, eye-tracking deficits with occlusion and agency attribution disorders were investigated. The circular inference model with E/I imbalance predicted findings in decision-making tasks involving likelihoods (e.g., Fisher task).
However, due to the conceptual design, their scalability is really poor for handling real sensory information. Here we find that ANNs, such as convolutional network approaches \cite{nagai2015influence} or Variational-Bayes RNN \cite{ahmadi2017bridging} could better account for real sensory data input.}

\annotation{Just as Hopfield networks were applied for modeling SZ, some early models of ASD focused on SOM approaches. These models \cite{gustafsson1997inadequate, gustafsson2004self, noriega2007self, noriega2008modeling} could account for strong specificity in cortical representations or novelty avoidance. Despite of that, they were highly linked to the specific network architecture, and thus, it is difficult to use these mechanisms to predict performance in other types of tasks.
More general approaches were suggested using simple parameter modifications of feed-forward neural networks \cite{cohen1994artificial, cohen1998neural, dovgopoly2013connectionist}. These parameters rather utilize general engineering mechanisms of neural networks and, thus, are also applicable to different architectures (e.g., regularization was also used in a recent approach using RNNs \cite{ahmadi2017bridging}). These studies mostly focused on replicating the specific symptom of generalization deficits, but may not be applicable to explaining a broader range of symptoms.}

\annotation{The reviewed models of SZ only addressed positive symptoms mainly hallucinations, delusions and abnormal movements. Self-other disturbances have been only discussed in the free-energy models and negative symptoms have been set aside. Within the ASD models only repetitive motor movements and hyper/hyporeactivity to sensory input were properly discussed. Furthermore, social communication and interaction deficits have been minimally addressed.}


\annotation{Interestingly, for ASD \cite{pellicano2012world, idei2017reduced, philippsen2018understanding, ahmadi2017bridging} as well as for SZ \cite{adams2013computational, jardri2013circular, yamashita2012spontaneous}, the majority of recent approaches incorporate the idea of predictive coding \cite{rao1999predictive}.
In particular, Pellicano and Burr's paper \cite{pellicano2012world} and novel hypotheses based on their theory \cite{lawson2014aberrant, lawson2017adults} significantly influence the recent developments.
In terms of finding a general account for cognition, predictive coding and related approaches are the most promising candidates right now. 
Therefore, predictive coding based approaches can be considered a useful abstraction in developing a broader model that is able to integrate typical and atypical development in a coherent whole.}

\subsection{Multifinality, equifinality and heterogenity}
\label{sec:discussion:multifinality}

\annotation{A challenge in modeling psychopathologies is the non-specificity of these disorders. Different biological bases may lead to the same symptom (equifinality). Therefore, many modeling mechanisms might be valid for modeling a single symptom. Accordingly, the studies reviewed here cover a wide range of approaches, using various pieces of biological evidence.
This variety has its drawback: even if a model can explain some symptoms, we can not judge whether this mechanism actually is comparable to what happens in the human brain or not.}

\annotation{The non-specificity of psychopathologies also means that a single biological basis can cause different symptoms (multifinality). Thus, instead of targeting single symptoms, it is important to develop models which explain several symptoms of a disorder. A good starting point is to first model typical behavior. One possible basis could be ANN models of sensorimotor integration. According to the majority of the computational models discussed in this manuscript, SZ and ASD are presented as disorders of sensory information fusion or interpretation. Thus, general ANN sensorimotor integration models that are able to fit human-like data (control and patient data) in different experimental paradigms such as body perceptual tests or decision making task could be extended to model psychopathologies.}

\annotation{Additionally, modeling mechanisms should not only cover various symptoms of a single disorder, but they may also be used for modeling similar symptoms in different disorders. For instance, hallucinations are present in several disorders but researchers used different ANN approaches to model them. Hallucinations produced by a loss of sensory input, like in the Charles Bonnet syndrome, were studied by modeling homeostasis in a Deep Boltzmann machine (DBM) for visual \cite{series2010hallucinations} and tactile inputs \cite{deistler2019tactileHallucinations}. However, homeostasis or DBMs were never studied for hallucinations in SZ, or discussed within circular inference or free-energy approaches \cite{adams2013computational}.}

\annotation{Regarding heterogeneity, recent studies modeling ASD already acknowledge the nature of ASD as a spectrum. Instead of distinguishing between impaired and intact behavior as two categories, a continuous change in symptoms is suggested, leading to impairments of different severeness \cite{nagai2019predictive, idei2017reduced} or even opposite types of impairments \cite{philippsen2018understanding}. This offers a potentially more sophisticated view on heterogeneity in ASD.}

\subsection{Models validation on real robotic systems}
\label{sec:discussion:realworld}

We presented some works that employ robotics systems' validation as a useful servant for the behaviour unit/level of analysis \cite{yamashita2012spontaneous}. The relevant aspect of these approaches is that the internal mechanism of the behaviour is visible \cite{cheng2007cb}.
\annotation{Furthermore, a connection can be made from rather perceptual or mechanistic impairments inside the system to difficulties in real interaction scenarios.} For instance, \cite{murata2013learning} replicated freezing and repetitive behaviors on a robot.
Most of the discussed models, however, are solely data models. Closing the gap to real world embodied models could, therefore, help to validate how these models extend to other tasks.

ANN approaches can also focus on solving scalability to raw stimuli in other brain-inspired mathematical abstractions. For instance, \cite{lanillos2018adaptive} and \cite{oliver2019active} presented free-energy-based perception and action algorithms working on humanoid robots. They can be used to evaluate atypical behaviours related to body perception in SZ and ASD.

\subsection{New directions}
\label{sec:discussion:newdirections}

\annotation{We identified the following three research directions that are still underrepresented in the discussed studies.}

\subsubsection{Developmental factors}
\label{sec:discussion:developmental}

Developmental factors are especially relevant for ASD as a developmental disorder, but also for SZ. Specially, to explain why many cases of SZ emerge during adolescence and early adulthood \cite{huttenlocher1979synaptic, feinberg1982schizophrenia, keshavan1994schizophrenia} and to investigate developmental factors which might contribute to the onset of SZ \cite{cannon2015schizophrenia}.
Current models only partially take the developmental process into account and focus more on modeling existing deficits in adult subjects with ASD.
For instance, existing models assume an aberrant number of neurons \cite{cohen1994artificial, noriega2007self} or differences in the neural connections \cite{ichinose2017local, park2019macroscopic} during the development, or they change the way that learning proceeds by altering network regularization \cite{dovgopoly2013connectionist, ahmadi2017bridging} or how information are integrated during learning \cite{philippsen2018understanding}.
However, these studies still cannot answer the question of which initial causes promote the appearance of ASD during the development.
It might be beneficial to take even one step more back in development, back to the development of the human fetus.
For instance, a recent study \cite{yamada2016embodied} suggests that disordered intrauterine embodied interaction during fetal period is a possible factor for neuro-developmental disorders like ASD.





\subsubsection{SZ and ASD as disorders of the self}
\label{sec:discussion:self}
One of the aspects not properly addressed in ANN computational modeling, neither for SZ nor for ASD, is how diagnosed individuals experience their body and self in comparison with control subjects. For instance, SZ patients have troubles differentiating self-produced actions. In fact, modeling the spectrum of differences in body experience could make several psychopathologies comparable. In addition to already described visual illusions, also body illusions can be investigated. Recently, Noel et al. \cite{noel2017spatial} discussed how body perception differs between ASD and SZ individuals, suggesting a sharper boundary between self and other in ASD and a weaker boundary in SZ. This suggestion is based on experimental findings, for example, on peripersonal space in body illusions where ``opposite" results were found: whereas individuals with SZ were more prone to have body illusions \cite{thakkar2011disturbances}, individuals with ASD showed a reduced illusionary effect \cite{cascio2012rubber}.
Hence, the causes of these psychopathologies have a direct impact on the perception of our body and the self. In the case of patients diagnosed with SZ, this relation has been more intensively studied \cite{stanghellini2009embodiment} and some treatments include embodiment therapies. Hence, models of the bodily or sensorimotor self \cite{lanillos2017enactive, hinz2018drifting} that are able to explain body illusions would help to validate the hypothesis in a common framework. Behavioural measures like the proprioceptive drift or peripersonal space should be also predicted by the model. For instance, in \cite{hinz2018drifting}, they used the perceptual drift as a measure to evaluate the validity of a predictive coding model for typical individuals.

\subsubsection{ANN novel architectures for psychopathologies}
\label{sec:discussion:ANN}
In terms of neural network architectures, there is a further need of transferring the knowledge from state-of-the-art recurrent neural networks and deep learning to neurological disorders as it was performed, for instance, with the Neocognitron model of ASD \cite{nagai2015influence} or the MTRNN model of SZ \cite{yamashita2012spontaneous}. Theoretical ANN studies, computational psychiatry and neuroscience should be always be in contact to boost the feedback of those disciplines. 

In opposition to Bayesian models that are implemented on a high abstraction level of the task, modern ANN approaches \cite{schmidhuber2015deep} are able to cope with real sensor data such as visual information. For instance, cross-modal learning architectures combined with hierarchical representation learning provide an interesting follow-up to early ANN studies on SZ and ASD. Furthermore, ANN models of Bayesian brain such as predictive coding \cite{yamashita2008emergence} and circular inference are a basis for uniting both communities. In fact, recent advances in probabilistic NNs like Variational Autoencoders \cite{kingma2013auto} and Variational-RNN \cite{fabius2014variational,ahmadi2017bridging}, provide the mathematical framework to deploy ANN versions of prominent plausible models of the brain such as the free-energy principle \cite{friston2010free}.

\annotation{In this review, we showed the power of ANNs for modeling symptoms of neurological disorders. However, these techniques need to be further developed and refined in the future to play a key role in computational psychiatry and to contribute in clinical applications.}

\section*{Acknowledgements}
This work was supported by SELFCEPTION project (www.selfception.eu) European Union Horizon 2020 Programme (MSCA-IF-2016) under grant agreement no. 741941, JST CREST Cognitive Mirroring (grant no. JPMJCR16E2), and by JSPS KAKENHI grant no. JP17H06039, JP18K07597 and JP18KT0021.

\section*{References}

\bibliography{references}

\end{document}